\documentclass[aps,pr,twocolumn,showpacs,superscriptaddress,groupedaddress,nofootinbib]{revtex4}  

\usepackage{longtable}
\usepackage{graphicx}  
\usepackage{ulem}   
\usepackage{comment} 
\usepackage{datetime}
\usepackage{dcolumn}

\usepackage{amsxtra}
\usepackage{euscript} 
\usepackage{bm}        
\usepackage{tabularx}
\usepackage{float}
\restylefloat{table}

\usepackage[cmtip,arrow]{xy}
\usepackage{pb-diagram,pb-xy}

\usepackage{amsmath}
\usepackage{amssymb}
\usepackage{amsthm}
\usepackage[pdftex]{color}
\usepackage[pdftex,colorlinks,citecolor=blue,linkcolor=blue,urlcolor=blue]{hyperref} 
\usepackage{graphicx}
\usepackage{dcolumn} 
\usepackage{bm} 

\usepackage{longtable}

\usepackage{ulem}   
\usepackage{comment} 
\usepackage{datetime}

\usepackage{tabularx}
\usepackage{float}
\restylefloat{table}

\newcommand{\beq}{\begin{equation}}
\newcommand{\eeq}{\end{equation}}
\newcommand{\bea}{\begin{eqnarray}}
\newcommand{\eea}{\end{eqnarray}}
\newcommand{\barr}{\begin{array}}
\newcommand{\earr}{\end{array}}

\long\def\begincomment#1\endcomment{}

\newcommand{\Tr}{\mathrm{Tr}}

\newtheorem{theorem}{Theorem}
\newtheorem{definition}{Definition}
\newtheorem{lemma}{Lemma}
\newtheorem{remark}{Remark}
\newtheorem{corollary}{Corollary}
\newtheorem{proposition}{Proposition}
\usepackage{graphicx}

\usepackage{pgf,tikz}
\usetikzlibrary{arrows}

\usepackage{bm}
\usepackage{bbold}

\usepackage{mathtools}

\DeclarePairedDelimiterX\braket[2]{\langle}{\rangle}{#1 \delimsize\vert #2}

\begin{document}


\title{Revisited functional renormalization group approach for random matrices in the large-$N$ limit}

\author{Vincent Lahoche} \email{vincent.lahoche@cea.fr}   
\affiliation{Commissariat à l'\'Energie Atomique (CEA, LIST),
 8 Avenue de la Vauve, 91120 Palaiseau, France}

\author{Dine Ousmane Samary}
\email{dine.ousmanesamary@cipma.uac.bj}
\affiliation{Commissariat à l'\'Energie Atomique (CEA, LIST),
 8 Avenue de la Vauve, 91120 Palaiseau, France}
\affiliation{International Chair in Mathematical Physics and Applications (ICMPA-UNESCO Chair), University of Abomey-Calavi,
072B.P.50, Cotonou, Republic of Benin}

\date{\today}

\begin{abstract}
\begin{center}
\textbf{Abstract}
\end{center}
The nonperturbative renormalization group has been considered as a solid framework to investigate fixed point and critical exponents for matrix and tensor models, expected to correspond with the so-called double scaling limit. In this paper, we focus on matrix models and address the question of the compatibility between the approximations used to solve the exact renormalization group equation and the modified Ward identities coming from the regulator. We show in particular that standard local potential approximation strongly violates the Ward identities, especially in the vicinity of the interacting fixed point. Extending the theory space including derivative couplings, we recover an interacting fixed point with a critical exponent not so far from the exact result, but with a nonzero value for derivative couplings, evoking a strong dependence concerning the regulator. Finally, we consider a modified regulator, allowing to keep the flow not so far from the ultralocal region and recover the results of the literature up to a slight improvement. \\

\noindent
\textbf{Key words :} Matrix models, tensor models, quantum gravity, random geometry.
\end{abstract}

\pacs{71.70.Ej, 02.40.Gh, 03.65.-w}

\maketitle
\section{Introduction}

Random matrix models are specific statistical models describing (Euclidean) quantum fluctuations of a matrixlike field \cite{DiFrancesco:1993cyw}. They appear as a framework for a very large kind of problems in physics and mathematics, from quantum gravity to biology (the list of references is very large, and we do not mention them here). In this paper, we essentially focus on quantum gravity interpretation, even if our results do not especially refer to this interpretation. The link between matrix models and two-dimensional quantum gravity arises from the observation that the perturbation series of random matrix models can generate randomly arbitrary triangulated surfaces (see \cite{Brezin:1992yc}-\cite{Ambjorn:1991cs} and references therein); the precise relation between Feynman diagrams and elementary polygons being discussed on a concrete example in section \ref{sec1}. Then, Feynman amplitudes of the perturbative expansion for such models are indexed by simplicial decomposition of the two-dimensional manifold; and as an important result (in particular for quantum gravity issues), the relative weight of two such a triangulation depends only on the genus of the corresponding manifold and the size $N$ of the considered matrices \cite{DiFrancesco:1993cyw}.

In the large $N$ limit, only planar surfaces survive, and the computation of the corresponding free energy shows the existence of a critical point, where infinitely refined triangulation starts to dominate the perturbative series; and interpreted as a continuum limit. Double scaling is a theoretical framework allowing keeping into account higher genus surfaces, taking the large $N$ limit near the critical point in such a way that the relative weight of the different topological configurations are exactly compensated by their growth, fixed by a universal critical exponent. Renormalization group (RG) has been considered to be an alternative to the standard analytic method \cite{Eichhorn:2013isa}-\cite{Sfondrini:2010zm}. The argument is based on the interpretation of the correlation between coupling and $N$ in the double scaling limit as a fixed point of the RG flow with $N$. In a Wilsonian perspective, integrating out matrix entries between $N$ and $N-\delta N$ generates effective actions, which drag the couplings so far from their initial values. This version of the RG flow and perturbative solutions has been investigated for twenty years \cite{Brezin:1992yc}-\cite{Pawlowski:2015mlf}, and reproduces semiquantitatively the exact results. More recently, a nonperturbative FRG framework has been considered to improve the perturbative results \cite{Eichhorn:2013isa}. In this reference paper, the authors show convergence phenomena for the computed critical exponents toward the exact (i.e. analytic result) for double scaling.

In the following paper, we show that the naive approaches to solve the nonperturbative RG equations, especially based on a reduction of the theory space to the strictly local interactions or products of them are strongly incompatible with Ward's identities \cite{Lahoche:2018ggd}-\cite{Takahashi:1957xn}. The origin of the incompatibility is traced to come from the regulator itself. Indeed, due to the presence of the regulator, the compatibility with Ward identities requires to enlarge the theory space to derivative couplings; which in turn seems to play a significant role in the fixed point structure, and finally introduce a spurious dependence on the regulator. To solve this issue, we introduce a modified regulator, parametrized in such a way that the contribution of derivative couplings in the Ward identities remains small in a significant domain of the RG flow, so that truncation involving only traces may be used without strong disagreements to approximate the exact solution of the RG equations.
Note that we explicitly checked that the method used to derive the flow equations in the reference \cite{Eichhorn:2013isa} (i.e. a systematic projection using a $U(N)$-invariant vacuum ansatz) is inconsistent, and the first part of this paper, voluntary pedagogical provide another derivation of the flow equations for truncation involving trace observables. \\

The outline is the following. Sections \ref{sec1} reviews shortly on the matrix models, double scaling and functional renormalization group approach, including all the preliminaries required for the rest of the discussion. In section \ref{sec3} we review the nonperturbative renormalization group flow following the reference \cite{Eichhorn:2013isa}, first in the local potential approximation and second including multitrace interactions. In section \ref{sec4} we show explicitly that the local potential approximation strongly violates Ward identities for some choice of natural regularization functions. We investigate the flow numerically and compare the numerical fixed point with the analytic solutions. In the last section \eqref{sec6} we provide some discussions and the conclusion of this work.

\section{Preliminaries}\label{sec1}
\subsection{A short review on matrix models}
\begin{figure}[h!]
$\underset{(a)}{\vcenter{\hbox{\includegraphics[scale=0.5]{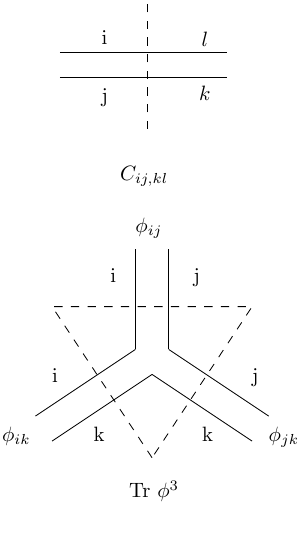} }}}\quad \underset{(b)}{\vcenter{\hbox{\includegraphics[scale=0.35]{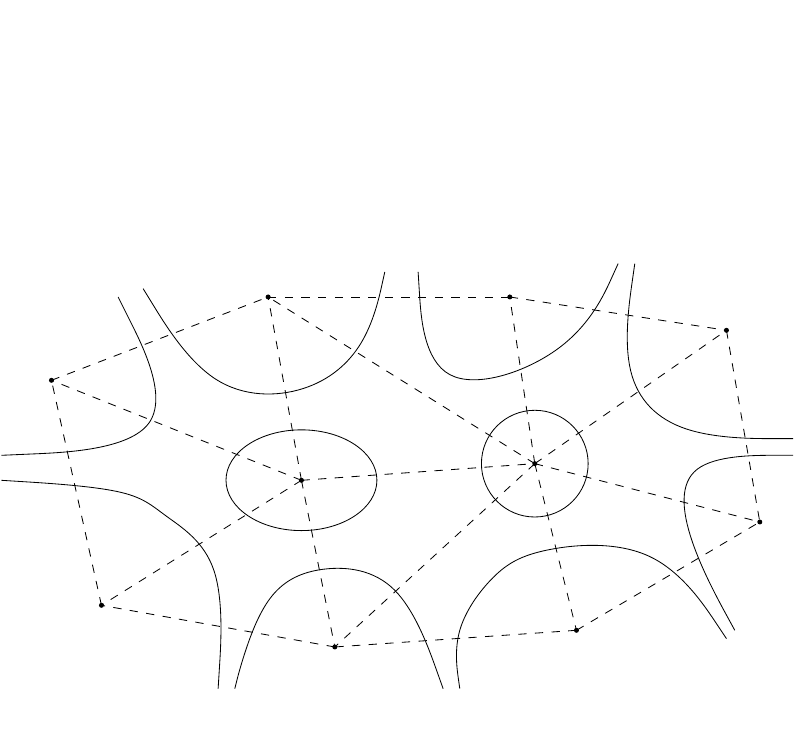} }}}$
\caption{Propagator and vertex of the trivalent matrix model. The dual representation is pictured with dotted edges: they correspond to an edge for a propagators, and to a triangle for a vertex (a). An example of a ribbon graph $\mathcal{G}$ and its corresponding dual triangulation $\Delta_{\mathcal{G}}$ (b).}\label{fig12}
\end{figure}

To shortly reviewing matrix models, let us consider a concrete example for a trivalent model involving Hermitian $N\times N$ matrix $\phi$, described from the partition function:
\begin{equation}
\mathcal{Z}:=\int d\phi e^{-\frac{1}{2} \Tr \phi^2-\frac{g}{\sqrt{N}} \Tr\phi^3}\,, \label{defmodel}
\end{equation}
where $dM$ is the invariant measure on the $N\times N$ Hermitian matrices (for more detail see \cite{Brezin:1992yc}-\cite{Zinn-Justin:2014wva}). The classical action $S[\phi]:=\frac{1}{2} \Tr \phi^2+\frac{g}{\sqrt{N}} \Tr\phi^3$ admits a natural $U(N)$ symmetry due to the global trace structure. Expanding the right hand side perturbatively in $\lambda$ with the propagator
\begin{equation}
C_{ij,kl}=\delta_{jk}\delta_{il}\ ,
\end{equation}
we generate Feynman amplitudes labeled by connected ribbon graphs $\mathcal{G}$, that is to say, a set of vertices, edges and faces. The interaction vertex has three external points, identifying the six strands pairwise. Propagator, vertex and their dual correspondence are pictured on Figure \ref{fig12}a, and an example of ribbon graph is given in Figure \ref{fig12}b.

From Feynman rules, the partition function expands as a sum of Feynman amplitudes
\begin{equation}
\mathcal{Z}=\sum_{\mathcal{G}}\frac{1}{s(\mathcal{G})}g^{n(\mathcal{G})}\mathcal{A}_{\mathcal{G}}\ , \label{eqmat}
\end{equation}
where, up to the rescaling $\phi\to\sqrt{N}\phi$, the amplitude $\mathcal{A}_{\mathcal{G}}$ depends on $N$ and on the genius $h$ of the dual representation $\Delta_{\mathcal{G}}$ of $\mathcal{G}$ : $\mathcal{A}_{\mathcal{G}}=N^{2-2h(\Delta_{\mathcal{G}})}$. We stressed that matrix models are statistical models for triangulated surfaces, but we have not made contact yet with quantum gravity in dimension two. This correspondence can be heuristically traced as follows. Including cosmological constant $\Lambda$, classical gravity in dimension two is described by the action:
\begin{equation}
\mathcal{S}_{2d}=\frac{1}{G}\int _{\mathcal{M}} d^2x\sqrt{-g}(-R(g)+\Lambda)=-\frac{4\pi}{G}\chi(\mathcal{M})+\frac{\Lambda}{G}A_{\mathcal{M}}\ , \label{2daction}
\end{equation}
where we used the Gauss-Bonnet theorem to compute the integral in terms of the Euler characteristic $\chi(\mathcal{M})$, and where we have denoted by $A_{\mathcal{M}}$ the area of the surface $\mathcal{M}$. Then, the theory only depends on two parameters, and we generally assume that only these two parameters are relevant to define the discretization. As a basic example, introducing an equilateral triangulation $\Delta_{\mathcal{M}}$ of $\mathcal{M}$, such that each triangle has a fixed area $a$, the action \eqref{2daction} can be discretized as
\begin{equation}
\mathcal{S}_{2d}(\Delta_{\mathcal{M}}):=-\frac{4\pi}{G}\chi(\Delta_{\mathcal{M}})+\frac{\Lambda a}{G}A_{\mathcal{M}}(\Delta_{\mathcal{M}})\ ,
\end{equation}
and the quantum theory described by the partition function:
\begin{equation}
\mathcal{Z}_{2d}=\sum_{\Delta} e^{\frac{4\pi}{G}\chi(\Delta_{\mathcal{M}})-\frac{\Lambda a}{G}A_{\mathcal{M}}(\Delta_{\mathcal{M}})} \label{discretized}
\end{equation}
matches the partition function \eqref{eqmat}, up to the identification:
\begin{equation}
g \leftrightarrow e^{-\Lambda a/G}\ ,\qquad N\leftrightarrow e^{4\pi/G}\,.
\end{equation}
As a result, heuristically, the large $N$ limit of matrix models (involving a lot of ``microscopic'' degrees of freedom) matches with the weak coupling regime of two dimensional topological gravity. Some formal results show the equivalence between matrix models and other quantum gravity approach. In particular, equivalence with Liouville theory at fixed topology has been stressed from agreement with KPZ relation, see \cite{Duplantier:2009np}-\cite{Duplantier:2009np2}. \\

The leading order contribution in the $N\to\infty$ limit comes from the triangulations with zero genus, corresponding to a planar topology. Interestingly, a closed two-dimensional topological manifold is fully characterized by its genus and orientability. Note that Hermitian matrices only generate orientable surfaces, so that the genus fully determines the topology of the triangulation and allows to capture nonperturbative effects. Indeed, the perturbative expansion \eqref{eqmat} can be rewritten as a topological expansion :
\begin{equation}
\mathcal{Z}=\sum_{h\in\mathbb{N}^*} N^{2-2h} \mathcal{Z}_h(g)\,,
\end{equation}
where we have defined the sum over all triangulations with genus $h$ as $\mathcal{Z}_h(g)$. In the large $N$ limit, the partition function for zero genus surfaces, $\mathcal{Z}_0(g)$ has the following critical behavior \cite{Brezin:1992yc}-\cite{Zinn-Justin:2014wva}:
\begin{equation}
\mathcal{Z}_0(g) \sim |g-g_c|^{2-\gamma}\ ,
\end{equation}
where $\gamma=-1/2$. As a result, the free energy of the planar sector diverges around the critical point $g=g_c$ corresponding to the continuum limit, where $\mathcal{Z}_0$ is dominated by arbitrary large triangulations. Going beyond the planar sector requires a \textit{double scaling limit}, taking the two limits $N\to \infty$ and $g\to g_c$ in a correlated manner \cite{DiFrancesco:1993cyw}. More precisely, we can show that $\mathcal{Z}_h$ has the same critical point as $\mathcal{Z}_0$ for any $h$:
\begin{equation}
\mathcal{Z}_h(g)\sim |g-g_c|^{\frac{(2-\gamma)(2-2h)}{2}}\,,
\end{equation}
suggesting to take simultaneously the limits $N\to \infty$ and $g\to g_c$ in such a way that the ratio
\begin{equation}
N|g-g_c|^{(2-\gamma)/2} \label{constraintds}
\end{equation}
remains fixed such that all the topologies contribute to the free energy when we are close to the critical point :
\begin{equation}
\mathcal{Z}\sim \sum_{h} f_h \big[N|g-g_c|^{(2-\gamma)/2}\big]^{2-2h}\,,
\end{equation}
corresponding to the continuum limit, where the area diverge, like for the critical behavior of the naive $N\to \infty$ limit. The double scaling limit may be analytically investigated, and exact results for $g_c$ and critical exponents have found, see \cite{DiFrancesco:1993cyw} for details. \\

\subsection{Flowing on the matrix theory space}\label{sec2}

In order to make contact with the reference papers \cite{Eichhorn:2013isa}-\cite{Eichhorn:2014xaa} in view to compare our results with the ones, and in contrast with the model considered in the previous section, we focus on a quartic model, describing a random Hermitian matrix with the classical action:
\begin{equation}
S[\phi]=\frac{1}{2} \Tr (\phi^2)+ \frac{g}{4} \Tr (\phi^4)\,.
\end{equation}
In addition to the $U(N)$ symmetry, this model as a discrete $\mathbb{Z}_2$ symmetry $\phi\to -\phi$, and generate squarulations rather than triangulations. This distinction is unimportant for the continuum limit that we will investigate, which does not depend on the choice of elementary discrete polygons used to build random surfaces. For this model, the critical value $g_c$ and the corresponding critical exponent $\theta$ in the continuum limit have been exactly computed \cite{DiFrancesco:1993cyw}:
\begin{equation}
g_c= -\frac{1}{12}\,,\qquad \theta=\frac{4}{5}\,. \label{exact}
\end{equation}
The elementary intuition allowing to consider renormalization group approach to investigate the continuum limit for matrix comes from the constraint \eqref{constraintds}, freely interpreted as a fixed point for an appropriate scaling in $N$, at which any change as $N\to N-\delta N$ may be compensated by a change $g\to g+\delta g$ of the coupling, without change of the continuum physics \cite{Brezin:1992yc}. This elementary observation suggests, in accordance with a Wilsonian point of view, to integrate out step by step the large $N$ degrees of freedom on lines and rows, of the $N\times N$ matrices, reducing them to $(N-1)\times (N-1)$ matrices after a single step, $(N-2)\times (N-2)$ matrices after two steps, and so on. To each step, $(N-i)\times (N-i)$ matrices are described by effective action, which is a sum of two pieces: The classical action for $(N-i)\times (N-i)$ matrices, and the fluctuations term arising from integration of $N-i+1$ degrees of freedom. As a result, to each step, the couplings have the discrete change rule:
\begin{equation}
g_{i+1}=g_{i}+\frac{1}{N} \beta(g_i)+\mathcal{O}(1/N)\,,
\end{equation}
where the notation suggests that we consider only the large $N$ limit to define the $\beta$-function. Computing $\beta(g_i)$ from a single step, we get the one-loop beta function \cite{Brezin:1992yc}:
\begin{equation}
\beta(g)\approx g+6g^2+\mathcal{O}(g^3)\,, \label{oneloop}
\end{equation}
which vanishes for $g_*=-1/6$, in qualitative accordance with the exact result \eqref{exact}; the string susceptibility $\gamma$ being related to the critical exponent $-\beta^\prime(g^*)$ as \cite{Brezin:1992yc}:
\begin{equation}
2-\gamma=-\frac{2}{\beta^\prime(g^*)}\,. \label{string}
\end{equation}

The accuracy may be explicitly improved taking into account higher couplings and loop effects, an observation which strongly motivates a nonperturbative analysis, as suggested in \cite{Eichhorn:2013isa}. In this reference, the authors introduced a FRG framework based on the Wetterich equation formalism. The new version of the Wilson RG procedure requires a splitting into modes, between UV scales (no fluctuations are integrated out) and IR scales (all the fluctuations are integrated out) dictating how the small distance fluctuations are integrated out. As we will see, different steps correspond to partial integration of modes between $N$ and $N-\delta N$, and following the standard strategy in FRG formalism, we introduce a new term in the classical action:
\begin{equation}
\Delta S_N [\phi]=\frac{1}{2}\sum_{a,b,c,d} \phi_{ab} [r_N(a,b) ]_{ab;cd} \phi_{cd}\,, \label{regulator}
\end{equation}
which behaves like a scale dependence mass term, the specific slicing in $N$ depending on the shape of the regulator $r_N(a,b)$. Among the standard properties of the regulator, we recall the following (for more explanations, the reader may be consult the standard reviews \cite{Delamotte:2007pf}:
\begin{enumerate}
\item $r_N(a,b)$ has to have a nonvanishing ‘‘infrared" limit, i.e. $r_N(a,b)\sim N$ for $(a+b)/2N\to 0$. \,
\item $r_{N}(a,b) \to 0$ in the ‘‘ultraviolet'' limit , i.e. $(a+b)/2N\to \infty$. \,
\item $r_{N}(a,b)$ has to vanish in the limit $N\to 0$, allowing to recover the original partition function.\,.
\item $r_{N}(a,b)$ has to be of order $\Lambda$ in the limit $N\to \Lambda$, $\Lambda$ referring to the size of the matrices.
\end{enumerate}
Introducing this mass term into the classical action, we replace the global description given by the referent generating functional $\mathcal{Z}[J]:=\int d\phi e^{-S[\phi]+ J\cdot\phi}$, by a one-parameter set of models $\{\mathcal{Z}_N[J]\}$ defined as:
\begin{equation}
\mathcal{Z}_N[J]:=\int\, d\phi\, e^{-S[\phi]-\Delta S_N [\phi]+J\cdot\phi}\label{part2}\,,
\end{equation}
where the dot product is defined as $A\cdot B:= \sum_{mn} A_{mn}B_{mn}$. Due to the scale dependence of the regulator, the long distance physics effects ($(m,n)\lesssim N$) acquire a large mass and are frozen out, whereas the small distance effects ($(m,n)>N$) are integrated out. The RG flow then relates $\mathcal{Z}_{N}$ to $\mathcal{Z}_{N-\delta N}$. The transcription of this relation goes through a first order differential equation:
\begin{equation}
\dot{\Gamma}_N=\frac{1}{2}\Tr \left[\dot{r}_N\left(\Gamma^{(2)}_N+r_N\right)^{-1}\right]\,,\label{Wett}
\end{equation}
which indicates how the average \textit{effective action} $\Gamma_N$ change in the windows of scale $[N,N-dN]$ -- the dot meaning derivative with respect to the RG parameter $t:=\ln N$: $\dot{X}=N\frac{d}{dN}X$. We recall that the average effective action is defined as slightly modified Legendre transform of the free energy $\mathcal{W}_N:=\ln\mathcal{Z}_N$ :
\begin{equation}
\Gamma_N[\Phi]+\Delta S_N[\Phi]=J\cdot \Phi-\mathcal{W}_N[J]\,,
\end{equation}
where $\Phi$ denotes the classical field:
\begin{equation}
\Phi_{mn}:= \frac{\partial \mathcal{W}_N}{\partial {J}_{mn}}\,.\label{classical}
\end{equation}
In the same way $\Gamma^{(2)}_N$ in equation \eqref{Wett} denotes the second derivative of the average effective action :
\begin{equation}
\left[\Gamma^{(2)}_{N}\right]_{mn;pq}:=\frac{\partial^2 \Gamma_k}{\partial \Phi_{mn}\partial {\Phi}_{pq}}\,.
\end{equation}
Even to close this section we have to add an important comment about the notion of canonical scaling. Scaling, that is to say the dependence of the quantities on the cutoff coming from their dimensions, plays generally an important role in renormalization. In standard quantum field theory for instance, dimensionality is closely related to renormalizability. For matrix models, the situation is quite different, because there are no referent space-time, no referent length and no canonical scaling coming from extra structure of the theory. However, the behavior of the RG flow with $N$ in the vicinity of the Gaussian fixed point (i.e. keeping only the part of the scaling which is independent of the couplings), provides an intrinsic notion of dimension, that we call \textit{canonical dimension}:
\begin{definition}
For any trace observable $g_k \Tr(\phi^k)$ in the classical action, the canonical dimension of the coupling constant $g_k$ is defined in the vicinity of the Gaussian fixed point as the part of the scaling in $N$ which is independent of $g_k$ and the other couplings.
\end{definition}
\noindent
We denote as $d_k$ the canonical dimension of $g_k$, so that the intrinsic scaling writes as $N^{d_k+\mathcal{O}(g_1,g_2,\,\cdots)}$. To find the explicit expression of $d_k$, we then have to be investigate the behavior of the Feynman diagrams with $N$. This may be traced from the link between two-dimensional quantum gravity recalled in the previous section. Up to the rescaling $\phi\to \sqrt{N} \phi$ we have stressed a relation between matrix coupling, $N$, Newton and cosmological constant. Keeping this relation implies that each Feynman diagrams scales exactly as $N^{2-2h}\equiv N^{\chi}$, where $\chi(\Delta):= V(\Delta)-E(\Delta)+F(\Delta)$ denote the Euler characteristic of the polygon decomposition $\Delta$, having $V$ vertices, $E$ edges and $F$ faces. It is not hard to see that this holds if and only if, up to the mentioned rescaling, the only $N$ dependence of the classical action comes from a global $N$ factor, enforcing the definition
\begin{equation}
d_k=-\frac{k-2}{2}\,,
\end{equation}
in agreement with formula \eqref{defmodel}. In this paper, we will consider also multitrace interactions at the level of the effective action, and we have to extend this formula for such  interactions. In order to remain in accordance with the expected scaling $N^\chi$, we impose to cancel the additional $N$ factors coming from the additional traces. As a result, for an observable of the form $\prod_{j=1}^n\Tr (\phi^{k(j)})$, one assigns the canonical dimension $d^{(j)}_{k(1),\cdots,k(j)}$:
\begin{equation}
d^{(j)}_{k(1),\cdots,k(j)}= d_{\sum_j k(j)}-(j-1) \,.
\end{equation}
For a double trace operator for instance $\Tr (\phi^k)\Tr(\phi^l)$ one gets $d_{kl}^{(2)}=-(k+l)/2$. As pointed out in \cite{Eichhorn:2013isa}, it is interesting to note that, even for a single trace operator, the canonical dimension is negative for $k>2$, meaning that all non-Gaussian couplings are irrelevant. In this situation, the improvement of the scaling coming from radiative corrections plays an essential role in the fixed point structure. \\

\subsection{Ward-Takahashi identity}

Ward-Takahashi identities are a general feature of symmetry in quantum field theory and may be viewed as a quantum version of the Noether's theorem, resulting in the translation invariance of the Lebesgue integration measure in the path integral definition of the partition function (the reader could consult \cite{Ward:1950xp}-\cite{Takahashi:1957xn} for the first derivation of these identities in QED). Their interest in RG investigations has been largely discussed in the literature, and more specifically in the context of tensorial field theories in \cite{Lahoche:2018ggd}-\cite{Takahashi:1957xn}. Our point of view is that Ward identities are nontrivial functional relations, depending on the regulator like flow equations, and with this respect have to take into account in the building of the RG approximations. This is what we will do in the next section. We will extend this discussion about the role of Ward identity in section \ref{sec3}. In complement, the reader may consult \cite{Lahoche:2019vzy}-\cite{Lahoche:2019cxt}. \\

\noindent
Without a  regulator term, only the source terms break the global $U(\Lambda)$ invariance for some cutoff $\Lambda$. Infinitesimal variations provide the identity:
\begin{equation}
\Gamma_{N, \bullet\cdots (ab)(ba)}^{(n)}=\Gamma_{N, \bullet\cdots (cb)(bc)}^{(n)}\,, \label{asymptot}
\end{equation}
to all orders of the perturbation. Note that for the rest of this paper we restrict our investigations  into the \textit{symmetric phase} where vanishing classical field $\Phi$ defined from equation \eqref{classical} is expected to be a good vacuum and all the odd correlation functions vanish identically. \\

\noindent
The regulator term $\frac{1}{2} \phi \,r_k \phi$ breaks explicitly the global $U(\Lambda)$ invariance, and adds a new contribution to the asymptotic Ward identity \eqref{asymptot}. Let us consider an infinitesimal unitary transformation $1+\epsilon$, $\epsilon$ being an infinitesimal anti- Hermitian operator. At the leading order, the transformation rule for the matrix field $\phi$ is:
\begin{equation}
\phi \to \phi^\prime = (1+\epsilon)\phi (1+\epsilon)^{\dagger}\approx \epsilon \phi-\phi\epsilon\,.
\end{equation}
At the leading order in $\epsilon$, the total variation of the generating functional $\mathcal{Z}_N$ writes as
\begin{equation}
\delta\mathcal{Z}_N= \int d\phi e^{-S_{N}[\phi,J]} \left[-\delta S[\phi]-\delta\Delta S_N[\phi]+\delta (J\cdot \phi)\right]\,. \label{var}
\end{equation}
Because $S$ is a sum of traces, $\delta S[\phi]=0$. The variation of the source term is noting but:
\begin{align}
\nonumber \delta (J\cdot \phi)= J\cdot \delta\phi&= \sum_{a,b,c} (J_{ab} \epsilon_{ac}\phi_{cb} -J_{ab} \phi_{ac}\epsilon_{cb} )\\
&=\sum_{a,b,c} (J_{ab} \phi_{cb} -J_{bc} \phi_{ba})\epsilon_{ac}\,.
\end{align}
The variation of the regulation term can be deduced in the same way:
\begin{align}
\delta\Delta S_N[\phi]= \sum_{a,b,c,d}\left[\delta\phi_{ab} [r_N(a,b) ]_{ab;cd} \phi_{cd} \right]\,,
\end{align}
where we assumed that $[r_N(a,b) ]_{ab;cd}=[r_N(a,b) ]_{cd;ab}$. This is exactly the same computation as for the source term, up to the replacement $J_{ab} \to \sum_{c,d}\,[r_N(a,b) ]_{ab;cd} \phi_{cd} $, leading to:
\begin{align}
\nonumber \delta\Delta S_N[\phi]=\sum_{a,b,c,d,e} &\phi_{de}\bigg[r_N(a,b) ]_{ab;de} \phi_{cb} \\
&\qquad-[r_N(c,b) ]_{bc;de} \phi_{ba}\bigg] \epsilon_{ac}\,.
\end{align}
Moreover we assumed that $r_N(a,b)$ is a symmetric function with respect to $a$ and $b$. Due to the translation invariance of the Lebesgue measure, $Z_N$ must be invariant up to a global reparametrization of the fields, therefore the variation of the left-hand side in \eqref{var} must be vanish $\delta \mathcal{Z}_N=0$. From the identity:
\begin{equation}
\int d\phi\, \phi_{ab} e^{-S_N[\phi,J]} = \int d\phi\, \frac{\partial}{\partial J_{ab}} e^{-S_N[\phi,J]}
\end{equation}
We finally deduce the following statement:
\begin{theorem}
\textbf{Ward-Takahashi identity.} In the symmetric phase, and along the path $N=\Lambda$ to $N=0$, the following relation holds:
\begin{align}
\nonumber \bigg\{ \frac{\partial}{\partial J_{de}}\bigg([r_N(a,b) ]_{ab;de} \frac{\partial}{\partial J_{cb}}-[r_N(c,b) ]_{bc;de} \frac{\partial}{\partial J_{ba}}\bigg)\\
- \left(J_{ab} \frac{\partial}{\partial J_{cb}} -J_{bc} \frac{\partial}{\partial J_{ba}}\right)\bigg\} e^{\mathcal{W}_N[J]}=0\,,\label{Ward}
\end{align}
where we adopted the Einstein convention for repeated indices. Note that there are no summation over indices $a$ and $c$.
\end{theorem}

\section{Solving the RG flow in the symmetric phase}\label{sec3}

The exact flow equation \eqref{Wett} cannot be solved exactly except for very special problems. Extracting information about the nonperturbative behavior of the RG flow then requires an appropriate scheme of approximation. In this section, we review a standard approach based on a crude truncation of the theory space. We focus on local interactions, or products of them, to remain closer to what we expect to be the theory space of the original matrix model, without regulator. As mentioned in the Introduction, this section is voluntarily pedagogical, due to strong disagreements with some results in the principal cited reference \cite{Eichhorn:2013isa}.

\subsection{Local potential approximation}

\textit{i.) Local potential.} The matrix action is nonlocal in the usual sense in field theory because all the interacting fields are not evaluated on the same point of the structure manifold. However, what allows to say that two objects interact locally is precisely the form of the interaction. The interaction then allows to define by themselves an appropriate locality principle, and we adopt the definition:
\begin{definition}
Any global trace of the form $\Tr (\phi^k)$ is said to be a local monomial interaction. In the same way, any functional of $U[\phi]$ which may be expanded as a sum of single traces is said to be a local functional.
\end{definition}\label{deflocal}
\noindent
Note that this locality principle reflects the proper invariance of the interactions concerning unitary transformations\footnote{See \cite{Lahoche:2019vzy}-\cite{Lahoche:2019cxt} for an extended discussion, showing how this definition works in practical contexts, especially in the context of matrix field theory to define counterterms for renormalization}. \\

\noindent
The first parametrization of the theory space that we consider split the effective action $\Gamma_N(\Phi)$ as a sum of two kind of terms:
\begin{equation}
\Gamma_N(\Phi)=(\text{nonlocal})+U_N(\Phi)\,. \label{decomp1}
\end{equation}
The last term $U_N(\Phi)$ designates the purely local potential, expanding as a sum of single trace observables:
\begin{equation}
U_N(\Phi)= \frac{Z_N}{2} \Tr (\Phi^2)+\frac{g_{4,N}}{4} \Tr (\Phi^4)+\frac{g_{6,N}}{6} \Tr(\Phi^6)+\cdots\,.\label{potential}
\end{equation}
Following \cite{Eichhorn:2013isa}, we introduced a field strength renormalization $Z_N$ in front of the Gaussian term. The renormalized quantities are generally defined from a fixed coefficient in the Gaussian part of the original action. Rescaling the fields such that the mass term reduces to its free term $\frac{1}{2} \Tr \Phi^2$, we define the dimensionless and renormalized couplings $u_{k,N}$ as:
\begin{equation}
u_{k,N}:= N^{-d_k} Z_N^{-k/2} g_{k,N}. \label{rencoupl}
\end{equation}
As pointed out in the derivation of the Ward identity, the presence of the regulator breaks the $U(\Lambda)$ invariance of the original action, and the RG flow has to generate noninvariante momentum dependent effective interactions such that, for instance:
\begin{equation}
K_N[\Phi]= \sum_{a,b} q\left(\frac{a}{N},\frac{b}{N}\right)\Phi_{ab}\Phi_{ba}\,. \label{kin}
\end{equation}
where the Taylor expansion of the function $q$ starts at the order $1$ in $a/N$ and $b/N$. The terminology ‘‘momentum dependent" simply reflects the situation in ordinary quantum field theory, the indices of the matrix field playing the role of discrete momenta. Expanding $q$ in power of $a/N$ and $b/N$ corresponds to the standard derivative expansion. As we will see from Ward identity, such a deviation from strict locality introduces relevant corrections at the leading order in $1/N$, and must be kept in the large $N$ limit. In particular, in the closure procedure around quartic interactions, the linear coupling:
\begin{equation}
q\left(\frac{a}{N},\frac{b}{N}\right)=\gamma\, \frac{a+b}{2N}\,,\label{nonlocal}
\end{equation}
plays an important role in the fixed point structure, improving strongly the local potential approximation. In a first time, in order to compare them, we keep only the strong local part of the decomposition \eqref{decomp1}:
\begin{equation}
\Gamma_N(\Phi)=U_N(\Phi)\,. \label{localpara}
\end{equation}

\begin{figure}
\includegraphics[scale=0.6]{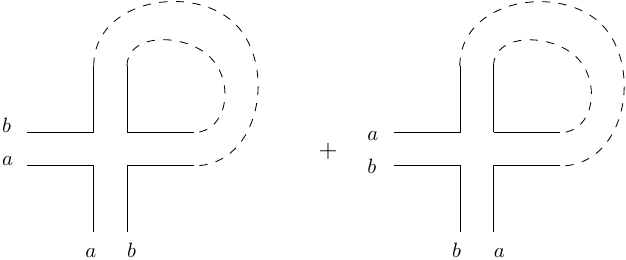}
\caption{Leading order contractions for 2-point graphs made of a single effective loop. } \label{figcont1}
\end{figure}

\noindent
The flow equations for couplings $g_n$ in the parametrization \eqref{localpara} can be deduced from the exact Wetterich equation deriving $n$ time with respect to $\Phi$ and setting $\Phi=0$ (we recall that we work in the symmetric phase). Because $\Phi$ is a Hermitian matrix, $\Phi_{ab}=\Phi^*_{ba}$, and :
\begin{equation}
\frac{\partial \Phi_{ab}}{\partial \Phi_{cd}}=\delta_{ac}\delta_{bd}\,,
\end{equation}
from which we get:
\begin{align}
[\Gamma_N^{(2)}]_{ab,cd}= \delta_{ac}\delta_{bd}Z_N\,,
\end{align}
where $g_{ab,cd}:=\delta_{ac}\delta_{bd}$ is nothing but the ‘‘bare" propagators. For the regulator function, we chose a modified version of the Litim optimized regulator \cite{Litim:2000ci}, allowing to make analytic computations:
\begin{equation}
\left[r_N(a,b)\right]_{ab,cd}=Z_N \delta_{cb}\delta_{ad} \left(\frac{2N}{a+b}-1\right) \Theta\left(1-\frac{a+b}{2N}\right)\,,
\end{equation}
which obviously satisfy the requirements $1-4$ given after equation \eqref{regulator}. Taking the derivative with respect to the flow parameter $t=\ln N$, we get straightforwardly:
\begin{align}
\nonumber \left[\dot{r}_N(a,b)\right]_{ab,cd}= Z_N g_{ba,cd}&\frac{2N}{a+b} \Theta\left(1-\frac{a+b}{2N}\right)\\
&+ \eta_N \left[r_N(a,b)\right]_{ab,cd}\,,
\end{align}
where we introduced the anomalous dimension
\begin{equation}
\eta_N:=\frac{ \dot{Z}_N}{Z_N}\,. \label{anmalous}
\end{equation}
Taking successive derivative with respect to $\Phi$ of the exact flow equation \eqref{Wett}, and setting $\Phi=0$, we deduce the flow equations for all couplings involved in \eqref{localpara}. For each step, all contributions involve some powers of the effective propagator $G_N=(\Gamma_N^{(2)}+r_N)^{-1}$, evaluated for vanishing $\Phi$, and for $a+b\leq 2N$ as:
\begin{align}
(G_N)_{ab,cd}= Z_N^{-1} g_{ba,cd}\, \frac{a+b}{2N} \,. \label{eqG}
\end{align}
The one-loop sums that we will encounter in the derivation of the flow equations are all of the form:
\begin{equation}
I_{a}^{(p)}:= \sum_b \left((G_N)_{ab,ba}\right)^p \left[\dot{r}_N(a,b)\right]_{ab,ba}\,.
\end{equation}
In the large $N$ limit, the sum can be replaced by an integration up to $1/N$ corrections. Let us introduce the continuous variable $2N x:=a+b$, running from $a/2N$ to $1$:
\begin{equation}
I_{a}^{(p)}\approx 2Z_N^{1-p} N \int_{a/2N}^{1} dx\,x^{p-1} \, \left(1+\eta_N(1-x)\right)\,, \label{Ip}
\end{equation}
leading to:
\begin{align}
\nonumber I_{a}^{(p)}\approx 2Z_N^{1-p} &N \bigg[\frac{1}{p} \left(1-\left(\frac{a}{2N}\right)^p\right)(1+\eta_N)\\
&\qquad-\eta_N \frac{1}{p+1} \left(1-\left(\frac{a}{2N}\right)^{p+1}\right)\bigg]\,.\label{intapprox}
\end{align}
\textit{ii.) Truncated RG flow.} Deriving the equation \eqref{Wett} twice with respect to the $\Phi$ fields, and setting $\Phi=0$, we get:
\begin{equation}
\dot{\Gamma}_{N,ab,ba}^{(2)}=-\frac{1}{2}\,G_{N,cd,ef} \Gamma^{(4)}_{N,ef,lm,ab,ba}\tilde{G}_{N,lm,cd}\,, \label{equationwettexp1}
\end{equation}
with $\tilde{G}_{N,ab,cd}:= ({G}_{N}\,\dot{r}_N)_{ab,cd} $ and where once again we sum over repeated indices. To compute the sums, we have to take into account the symmetry structure of the external indices. From \eqref{eqG}, we get for instance
\begin{equation}
G_{N,cd,ef}({G}_{N}{r}_N)_{lm,cd}= \frac{g_{ef,lm}}{Z_N}\, \left(\frac{l+m}{2N} \right)^2f(l/N,m/N)\,,\label{eq52}
\end{equation}
where we defined:
\begin{equation}
f(a/N,b/N)=\left(\frac{2N}{a+b}-1\right) \Theta\left(1-\frac{a+b}{2N}\right)\,.
\end{equation}
For a fixed configuration of the external indices, there are two leading order contractions, both pictured on Figure \ref{figcont1}, where in this graphical representation the dotted edges correspond to the contraction with the effective propagator $G_{N,cd,ef}\tilde{G}_{N,lm,cd}$ given by equation \eqref{eq52}. \\

\begin{figure}
\includegraphics[scale=0.7]{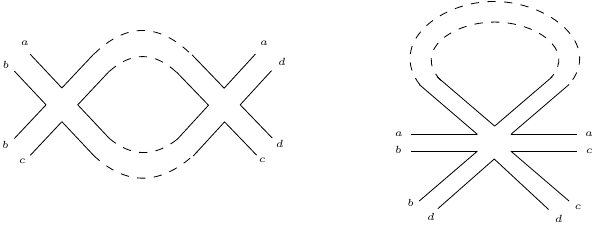}
\caption{Two typical leading order contractions contributing to the flow equation for $g_4$.} \label{fig2}
\end{figure}

\noindent
We now have to compute how many leading order contractions such as the one pictured on Figure \ref{figcont1} contribute. It is not hard to see that there are exactly $4\times 2$ different ways to build such a diagram : four different positions for the first end point of the propagator edge, and two remaining positions for the second end point, corresponding to the two attributions for the two free external edges, sharing the momentum $(a,b)$. Then, translating the diagram into equation, and setting $a=b=0$, one gets using the integral approximation \eqref{intapprox}
\begin{equation}
\vcenter{\hbox{\includegraphics[scale=0.4]{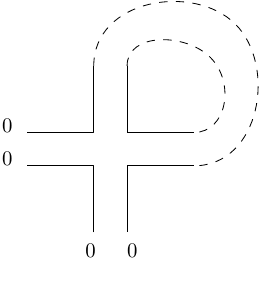} }}=2g_{4,N}I_0^{(2)}\,.
\end{equation}
Finally, computing the derivative of the left-hand side of equation \label{equationwettexp} for zero external momenta, we get :
\begin{equation}
{\Gamma}_{N,00,00}^{(2,0)}=Z_N\,,
\end{equation}
from which we deduce that:
\begin{equation}
\dot{Z}_N=-\frac{2Ng_{4,N}}{Z_N}\left(\frac{1}{2}+\frac{\eta_N}{6}\right)\,.\label{eqZ}
\end{equation}
Divided by $Z_N$, and from definitions \eqref{rencoupl}, we then get finally;
\begin{equation}
\eta_N=-\dfrac{3u_{4,N}}{3+u_{4,N}}\,.
\end{equation}
The computation of the beta function $\beta_4:=\dot{u}_{4,N}$ follows the same strategy. Deriving once again twice with respect to the $\Phi$ fields, and setting $\Phi=0$ at the end of the computation, one gets, formally:
\begin{equation}
\dot{\Gamma}_{N}^{(4)}=3\Tr\,\tilde{G}\,\Gamma_N^{(4)}G\,\Gamma_N^{(4)}G-\frac{1}{2}\Tr\,\tilde{G}\,\Gamma_N^{(6)} G\,.
\end{equation}
The relevant diagrams corresponding to the two kinds of traces involved in these expressions, all including one internal face are pictured on Figure \ref{fig2}.
Each of them may be easily translated into an equation like for the $2$-point diagrams. For zero external momenta we get:
\begin{equation}
\vcenter{\hbox{\includegraphics[scale=0.5]{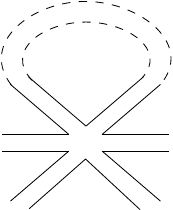} }}= 24\times  g_{6,N} I_0^{(2)}\,,
\end{equation}
and:
\begin{equation}
\vcenter{\hbox{\includegraphics[scale=0.5]{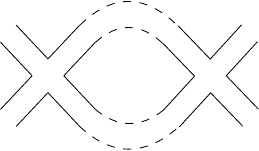} }}= 8\times g_{4,N}^2 I_0^{(3)}\,.
\end{equation}
Once again the numerical factors may be easily understood. For instance, for the diagram involving a $6$-point vertex, there are six different ways to choose the first end point of the contracted edge, two to choose the second one, to make a leading order graph; and finally $4!$ ways to exchange the remaining external points. Because from definition ${\Gamma}_{N,00,00,00,00}^{(4)}=6g_{4,N}$, it follows that:
\begin{equation}
6\dot{g}_{4,N}=24 g_{4,N}^2 I_0^{(3)} - 12 g_{6,N} I_0^{(2)}\,,
\end{equation}
leading to:
\begin{equation}
\beta_4=(1-2\eta_N) u_{4,N} + \frac{4u_{4,N}^2}{6}(4+\eta_N)-2u_{6,N}\left(1+\frac{\eta_N}{3}\right)\,. \label{eqbeta4}
\end{equation}
\begin{remark}
Neglecting the coupling $u_{6,N}$ and expending the remaining right-hand side in power of $u_{4,N}$, up to order $u^{3}_{4,N}$, we do not reproduce the one-loop result \eqref{oneloop}. In particular, the numerical factor in front of $u_{4,N}$ becomes $14/3$. This cannot be viewed as a defect of the approach, the $1$-loop beta function being nonuniversal for coupling with nonzero canonical dimension, as it can be easily checked.
\end{remark}

\noindent
Following the same procedure we can compute beta function for higher couplings, the flow equation for $g_{k,N}$ involving $g_{k+2,N}$ and so on, providing an infinite tower of hierarchical equations. The truncation method is the simpler approximation procedure, which truncates crudely in the full theory space, setting $g_{k,N}\approx 0$ for some $k$. This method has the advantage to be very tractable for (strict) nonlocal interactions, which is the case for matrix models. For $k=8$, i.e. setting $g_{8,N}\approx 0$ we find for the coupling $g_{6,N}$:
\begin{equation*}
\dot{\Gamma}_{N}^{(6)}=15\Tr \,\tilde{G}\,\Gamma_{N}^{(6)}G\,\Gamma_N^{(4)}G-45\,\Tr\, \tilde{G}\,\Gamma_N^{(4)}\, G\, \Gamma_N^{(4)}G\,\Gamma_N^{(4)}G\,,
\end{equation*}
where we neglected the term $-\frac{1}{2} \Tr \,\tilde{G} \Gamma_N^{(8)} G$. Computing each trace like for the two previous cases, we get two relevant diagrams at leading order:
\begin{equation}
5! \dot{u}_6\sim 15\left(\vcenter{\hbox{\includegraphics[scale=0.7]{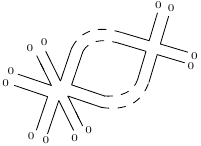} }}\right)-45\left(\vcenter{\hbox{\includegraphics[scale=0.7]{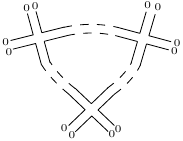} }}\right) \,.
\end{equation}
Taking into account the permutation symmetries the two diagrams are respectively evaluated to $4!\times 4g_6g_4 I^{(3)}_0$ and $2^4 g_4^3 I^{(4)}_0$; and we obtain for $\beta_{6}$:
\begin{align}
\nonumber \beta_6=2u_{6,N}-3\eta_N u_{6,N}&+2u_{6,N}u_{4,N} (4+\eta_N)\\
&-3u_{4,N}^3 \left( 1+\frac{\eta_N}{5}\right)\,.
\end{align}
To summarize, we have the following statement:
\begin{proposition}\label{prop1}
In the large $N$ limit, and in the local potential approximation, the truncated RG flow around $\phi^6$ interactions is described by the following closed system:
\begin{equation*}
\beta_4=(1-2\eta_N) u_{4,N} + \frac{2u_{4,N}^2}{3}(4+\eta_N)-2u_{6,N}\left(1+\frac{\eta_N}{3}\right)\,,
\end{equation*}
\begin{align*}
\nonumber \beta_6=(2-3\eta_N+2u_{4,N} (4+\eta_N)) u_{6,N}-3u_{4,N}^3 \left( 1+\frac{\eta_N}{5}\right)\,,
\end{align*}
with:
\begin{equation*}
\eta_N=-\dfrac{3u_{4,N}}{3+u_{4,N}}\,.
\end{equation*}
\end{proposition}
Note that the truncated RG flow becomes singular for $u_{4,N}=-3$, splitting the reduced phase space into disconnected regions. We call the perturbative region the region connected to the Gaussian fixed point. Moreover, we can remark that this singularity holds for arbitrary higher truncations. These equations allow to investigate the existence of nontrivial interacting fixed points for quartic and sextic truncations. \\

\begin{figure}
\includegraphics[scale=0.5]{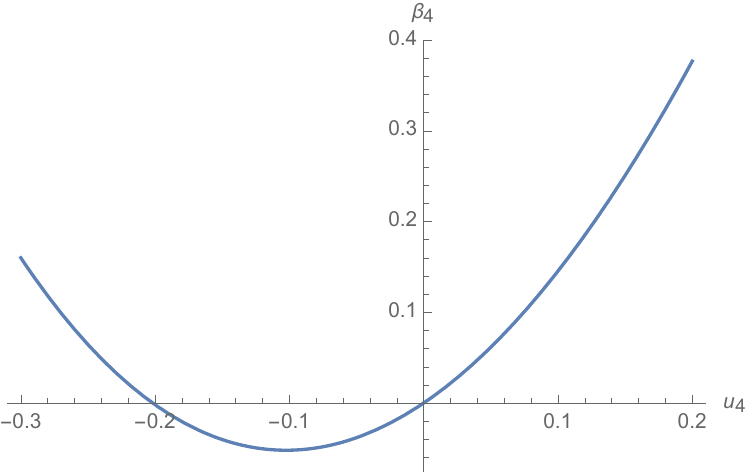}
\caption{Numerical plot of the beta function $\beta_4$. Except the Gaussian fixed point, we get only one UV-attractive interactive fixed point, for the value $u_4\approx -0.2$.} \label{figplot1}
\end{figure}

\noindent
$\bullet$ For $k=6$, the fixed point equation reduces to:
\begin{equation}
\beta_4=(1-2\eta_N) u_{4,N} +\frac{2}{3} u_{4,N}^2(4+\eta_N)=0\,.
\end{equation}
The numerical plot of the beta function is given in Figure \ref{figplot1}. We get two solutions:
\begin{equation}
u_{4,*}\approx -22.3\,,\quad \text{and} \quad u_{4,*}\approx -0.20 \,.
\end{equation}
The first solution is under the singularity line $u_{4,N}=-3$, and therefore unconnected to the Gaussian fixed point. The second solution however is in the perturbative region, and corresponds to an UV-attractive fixed point. Computing the anomalous dimension and the critical exponents, we get: $\eta_*\approx 0.21$ and $\theta_*=1.06$.\footnote {We recall that the critical exponents are the opposite of the eigenvalues of the matrix with entries $\partial_{u_{i}} \beta_j$.} \\

\noindent
$\bullet$ For $k=8$, the flow equations are given by the Proposition \ref{prop1}. Solving numerically the two equations $\beta_4=\beta_6=0$, we get once again two nontrivial interacting fixed point, for coordinates:
\begin{equation}
p_1:=(u_{4,*}^{(1)},u_{6,*}^{(1)})\approx (-0.27,0.05)\,,
\end{equation}
and
\begin{equation}
p_2:=(u_{4,*}^{(2)},u_{6,*}^{(2)})\approx (-0.14,-0.02)\,,
\end{equation}
with anomalous dimensions respectively given by $\eta_1\approx 0.3$ and $\eta_2\approx 0.15$ and critical exponents:
\begin{equation*}
(\theta^{(1)}_1,\theta^{(1)}_2)\approx (1.09,2.13)\,,\quad (\theta^{(2)}_1,\theta^{(2)}_2)\approx (1.04,-1.03)\,.
\end{equation*}
As expected, the result seems to be improved when the order of the truncation is increased. The fixed point that we found is reminiscent to the standard Wilson-Fisher fixed point, with only one attractive and one repulsive direction in the UV (i.e. in the large $N$ limit); the single positive critical exponents having to play the role of $\beta^\prime(g^*)$ in equation \eqref{string}.\\

The reliability of these results may be traced by investigating higher truncations. For $k=10$, we have to add the contribution $-\frac{1}{2} \Tr \,\tilde{G} \Gamma_N^{(8)} G$ for $\beta_6$,
\begin{equation}
\vcenter{\hbox{\includegraphics[scale=0.7]{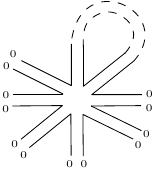} }}=\times 6!\,g_8\,I_0^{(2)}\,,
\end{equation}
which becomes:
\begin{align*}
 \beta_6=(2-3\eta_N+2u_{4,N} (4+\eta_N)) u_{6,N}&-3u_{4,N}^3 \left( 1+\frac{\eta_N}{5}\right)\\
&-u_{8,N}(3+\eta_N)\,.
\end{align*}
For $\dot{u}_8$, taking into account only the leading order contractions, we get:
\begin{align*}
7!\,& \dot{u}_8 = 28\,\left(\vcenter{\hbox{\includegraphics[scale=0.7]{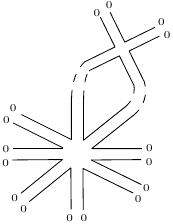} }}\right)-630\,\left(\vcenter{\hbox{\includegraphics[scale=0.7]{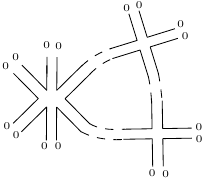} }}\right)\\
&+35\,\left(\vcenter{\hbox{\includegraphics[scale=0.7]{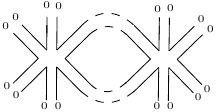} }}\right)+1260\,\left(\vcenter{\hbox{\includegraphics[scale=0.7]{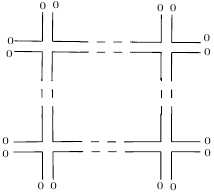} }}\right)\,
\end{align*}
from which we deduce for $\beta_8$:
\begin{align*}
\beta_8=(3&-4\eta_N)u_{8,N}+\frac{8}{15}u_{4,N}^4 (6+\eta_N)+\frac{4}{3}u_{6,N}^2\left(4+\eta_N\right)\\
&-12u_{6,N}u_{4,N}^2\left(1+\frac{\eta_N}{5}\right)+\frac{8}{3}u_{8,N}u_{4,N} \left(4+\eta_N\right)\,.
\end{align*}
Solving numerically the flow  equations, we get three fixed points, the first one being
\begin{equation}
(u_{4,*},u_{6,*},u_{8,*})\approx (-0.10, -0.025, -0.005)\,,
\end{equation}
with critical exponents:
\begin{equation}
(\theta_1,\theta_2,\theta_3)=(1.03,-0.97,-2.19)\,,
\end{equation}
and anomalous dimension $\eta_* \approx 0.11$. \\

Once again we find some results in qualitative accordance with the exact computation. We recover a Wilson-Fisher like fixed point having the expected characteristics, only one relevant direction with positive critical exponent. However, we do not observe significant improvement concerning $k=8$ truncation. This seems to indicate that higher irrelevant operators do not contribute much to the accuracy of the critical exponents. Note that this result is in complete disagreement with the ones of \cite{Eichhorn:2013isa}, where a convergence phenomenon has been pointed out by the authors. We suspect that this disagreement is a consequence of the method used by the authors, which, setting a diagonal vacuum $\Phi_{ab}=a\delta_{ab}$ to extract the flow equations, and therefore have selected more than strictly local interactions. \\

Interestingly, the numerical critical value for the coupling seems to be so far from the exact values than the ones obtained from $k=6$ and $k=8$ truncations. This value is not universal so that a disagreement with the exact value cannot be relevant for the reliability of the analysis. One expects that this is a defect of the LPA. Indeed, the fixed point arises essentially from the flow of irrelevant operators, which may be strongly coupled at the fixed point, where irrelevant interactions for the Gaussian counting can contribute significantly. Then, when we take into account higher interactions in LPA, we lost more and more information, coming especially from nonlocal and multitrace operators, as pointed out in \cite{Eichhorn:2013isa}. To investigate the improvement coming from these operators, let us consider the $k=8$ truncation, involving double and triple traces (to simplify the notation, we left the $N$ index for couplings):
\begin{align}
\nonumber \Gamma_N[\Phi]=&\frac{Z}{2} \Tr (\Phi^2)+\frac{g_4}{4}\Tr(\Phi^4)+ \frac{g_6}{6} \Tr(\Phi^6) + \frac{h_{2,2}}{4} (\Tr(\Phi^2))^2\\
&\qquad+\frac{h_{4,2}}{2} \Tr(\Phi^2)\Tr(\Phi^4)+\frac{h_{2,2,2}}{6} (\Tr(\Phi^2))^3+\cdots
\end{align}
The truncation for local interaction was based on the canonical dimension. For $k=8$ for instance, one can say that we discarded interactions with a canonical dimension smaller than $d=-3$. If we think to build the same truncation including multitrace interactions, we could conclude that interactions such that $\Tr(\Phi^2)\Tr(\Phi^4),$ which have canonical dimension $d=-3$ must be discarded like $\Tr(\Phi^8)$ interactions. However, the double trace increases the strength of the coupling. Then, in contrast to $\Tr(\Phi^8)$, an interaction such that $\Tr(\Phi^2)\Tr(\Phi^4)$ contributes directly to the flow of $g_4$ at leading order, the tadpole contraction scaling as $N^2$. \\

\noindent
Starting with the computation of $\dot{Z}$, we show that the contribution \eqref{eqZ} holds, but has to be completed with double-trace diagrams. Then,at the leading order in $N$, we get, graphically:
\begin{equation}
\dot{Z}=-\vcenter{\hbox{\includegraphics[scale=0.5]{contraction2.pdf} }}-\, \vcenter{\hbox{\includegraphics[scale=0.5]{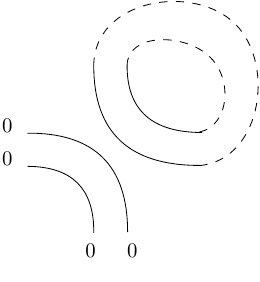} }}\,.
\end{equation}
Computing the new diagram, and taking into account that we have $4\times 2$ different permutations leading to the same diagram, we obtain:
\begin{equation}
\vcenter{\hbox{\includegraphics[scale=0.4]{contraction2bis.pdf} }}=\frac{1}{2} h_{22} J^{(2)}\,,
\end{equation}
where we defined $J^{(p)}$ as:
\begin{equation}
J^{(p)}:= \sum_{a,b} \left((G_N)_{ab,ba}\right)^p \left[\dot{r}_N(a,b)\right]_{ab,ba}\,,
\end{equation}
which can be approached by an integral as
\begin{equation}
J^{(p)}\approx 4N^2 Z^{1-p}\frac{2+p+\eta_N}{2+3p+p^2}\,.
\end{equation}
Finally, defining the dimensionless and renormalized couplings $v_{i,j,k,\cdots}$ as:
\begin{equation}
v_{i,j,k,\cdots}=N^{-d_{i,j,k,\cdots}}(\sqrt{Z})^{-i-j-k-\cdots} h_{i,j,k,\cdots}\,,
\end{equation}
we obtain, in replacement of the equation \eqref{eqZ}:
\begin{equation}
\eta=-u_{4} \left(1+\frac{\eta}{3}\right)-\frac{1}{3}v_{2,2} \left(4+\eta\right)\,,
\end{equation}
leading to:
\begin{equation}
\eta=-\frac{3u_4+4v_{2,2}}{3+u_4+v_{2,2}}\,.
\end{equation}
In the same way, for $\beta_4$, the previous computation has to be completed with the diagram:
\begin{equation}
\vcenter{\hbox{\includegraphics[scale=0.5]{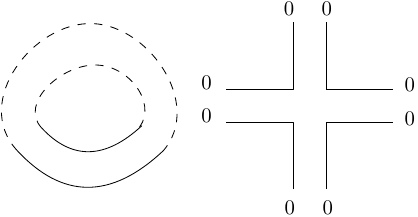}}} = 4!\,h_{4,2} J^{(2)}\,,
\end{equation}
leading to:
\begin{align*}
\beta_4=(1-2\eta) u_{4} + \frac{2u_{4}^2}{3}(4+\eta)-2u_{6}\left(1+\frac{\eta}{3}\right)-2v_{4,2}\frac{4+\eta}{3}\,.
\end{align*}
Finally, at this order for the truncation, the expression for $\beta_6$ is unaffected the multitrace interactions, except through the improvement of $\eta$. Now, we move on to the computation of the remaining beta functions, $\beta_{2,2}$, $\beta_{4,2}$ and $\beta_{2,2,2}$, respectively for couplings $v_{2,2}$, $v_{4,2}$ and $v_{2,2,2}$. For $\beta_{2,2}$, we get two kinds of leading order contractions, nonvanishing ones being:
\begin{align}
\nonumber 3! \dot{h}_{22}\sim &\vcenter{\hbox{\includegraphics[scale=0.55]{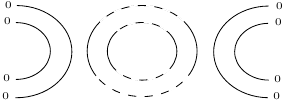}}}+\vcenter{\hbox{\includegraphics[scale=0.55]{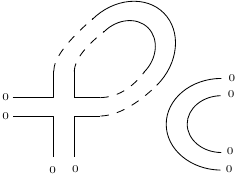}}}\\
&\qquad -\vcenter{\hbox{\includegraphics[scale=0.55]{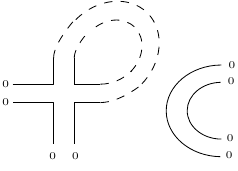}}}-\vcenter{\hbox{\includegraphics[scale=0.7]{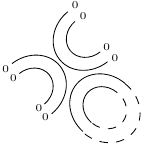}}}\,.
\end{align}
For the first kind of diagram, we get the contribution:
\begin{equation}
\vcenter{\hbox{\includegraphics[scale=0.5]{contraction4bis.pdf}}} = 6 h_{22}^2 J^{(3)}\,.
\end{equation}
The second kind of diagrams comes from the term $3\Tr\,\tilde{G}\,\Gamma_N^{(4)}G\,\Gamma_N^{(4)}G$, with interactions $(\Tr(\Phi^2))^2$ and $\Tr(\Phi^4)$:
\begin{equation}
\vcenter{\hbox{\includegraphics[scale=0.6]{contraction5bis.pdf}}}=12 g_4h_{22} I_0^{(3)}\,.
\end{equation}
Finally, the third and last contribution involving a nontrivial loop arises from the term $\frac{1}{2}\Tr\,\tilde{G}\,\Gamma_N^{(6)} G$, with interaction $\Tr(\Phi^2)^3$:
\begin{equation}
\vcenter{\hbox{\includegraphics[scale=0.8]{contraction6bis.pdf}}}=6 h_{222} J^{(2)}\,.
\end{equation}
Note that contributions involving loop without sum, such that the external indices fix the momentum along the loop vanish identically for zero external momenta. Indeed:
\begin{equation}
\left((G_N)_{ab,ba}\right)^p \left[\dot{r}_N(a,b)\right]_{ab,ba}\bigg\vert_{a=b=0}=0\,,\quad \forall\,p>1\,.
\end{equation}
Therefore, we obtain for $\beta_{22}$:
\begin{align}
\nonumber \beta_{22}=(2-2\eta)v_{22}+&2 v_{22}^2 \frac{5+\eta}{5}+\frac{4}{3}u_4v_{22} (4+\eta)\\
&-\frac{4}{3} v_{42} \left(1+\frac{\eta}{3}\right)- 2 v_{222} \frac{4+\eta}{3}\,.
\end{align}
In the same way, for $\beta_{4,2}$ and $\beta_{2,2,2}$, the nonvanishing typical diagrams are the following:
\begin{align}
\nonumber\dot{h}_{4,2}&\sim \vcenter{\hbox{\includegraphics[scale=0.7]{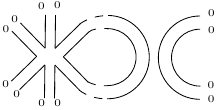}}}+\vcenter{\hbox{\includegraphics[scale=0.6]{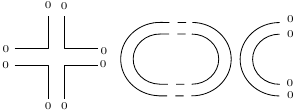}}}\\\nonumber
&+\vcenter{\hbox{\includegraphics[scale=0.6]{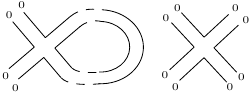}}}+\vcenter{\hbox{\includegraphics[scale=0.6]{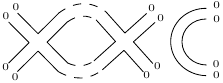}}}\\
&-\vcenter{\hbox{\includegraphics[scale=0.6]{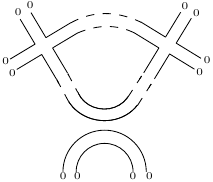}}}
\end{align}
and:
\begin{align}
\nonumber& \dot{h}_{2,2,2}\sim \vcenter{\hbox{\includegraphics[scale=0.7]{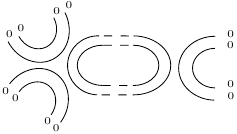}}}+ \vcenter{\hbox{\includegraphics[scale=0.7]{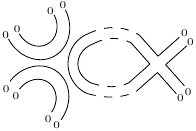}}}\\
& + \vcenter{\hbox{\includegraphics[scale=0.7]{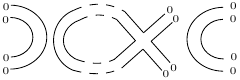}}}- \vcenter{\hbox{\includegraphics[scale=0.7]{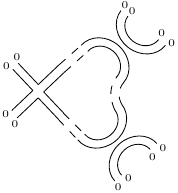}}}-\vcenter{\hbox{\includegraphics[scale=0.7]{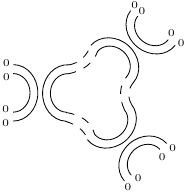}}}\,,
\end{align}
leading to the following statement:
\begin{proposition}
In the large $N$ limit, the multitrace truncated RG flow around sixtic interactions interaction is described by the following set of equations:
\begin{align*}
\beta_4=(1-2\eta) u_{4} + \frac{2u_{4}^2}{3}(4+\eta)-2u_{6}\left(1+\frac{\eta}{3}\right)-2v_{4,2}\frac{4+\eta}{3}\,,
\end{align*}
\begin{align*}
\nonumber \beta_{22}=\left(2-2\eta+ v_{22} \frac{5+\eta}{5}\right) v_{22}&+\frac{4}{3}u_4v_{22} (4+\eta)\\
&\qquad\quad - 2 v_{222} \frac{4+\eta}{3}\,,
\end{align*}
\begin{align*}
\beta_{42}=\bigg(3-3\eta&+2v_{22}\frac{5+\eta}{5} +\frac{4}{3}u_4 (4+\eta) \bigg)v_{4,2} \\
&\qquad \quad+\frac{2}{3} u_6v_{22} (4+\eta)-3v_{22} u_4^2 \left(1+\frac{\eta}{5}\right)\,,
\end{align*}
\begin{align*}
\beta_{222}=\bigg(4-3\eta+&6v_{22} \frac{5+\eta}{5}+2u_4(4+\eta)\bigg)v_{222}\\
&+\frac{1}{3} v_{42} v_{22} (4+\eta)- 36 u_4 v_{22}^2 \left(1+\frac{\eta}{5}\right)\\
&-\frac{16}{5}\,(v_{22})^3 (6+\eta)\,,
\end{align*}
\begin{align*}
\nonumber \beta_6=(2-3\eta+2u_{4} (4+\eta)) u_{6}-3u_{4}^3 \left( 1+\frac{\eta}{5}\right)\,,
\end{align*}
with:
\begin{equation}
\eta=-\frac{3u_4+4v_{2,2}}{3+u_4+v_{2,2}}\,.
\end{equation}
\end{proposition}

\noindent
As for the single trace potential, we may investigate numerically the fixed point structure, increasing progressively the degree of the truncation. \\

\noindent
$\bullet$ For $k=6$, the set of equations that we have to solve is the following :
\begin{align*}
\beta_4=(1-2\eta) u_{4} + \frac{2u_{4}^2}{3}(4+\eta)=0\,,
\end{align*}
and
\begin{align*}
\beta_{22}=2\left(1-\eta+ v_{22} \frac{5+\eta}{5}\right) v_{22}&+\frac{2}{3}u_4v_{22} (4+\eta)=0\,.
\end{align*}
Numerically, we get only one interesting fixed point, for $v_{22}=0$ and $u_4\approx -0.20$. This fixed point corresponds to the one discovered from the single trace $k=6$ truncation, therefore, we expect that there are no improvements coming from multitrace at this order of approximation. \\

\noindent
$\bullet$ For $k=8$, we may distinguish two cases. In the first one we only consider the effect of double-trace interactions, neglecting the triple trace. In the second one, we include the triple-trace interaction. For the double-trace approximation, we get only one potential candidate fixed point, for the values:
\begin{equation}
(u_{4*},v_{22*},u_{6*},v_{42*})\approx (-0.14,0,-0.02,0)\,,
\end{equation}
with anomalous dimension $\eta_*\approx 0.14$ and critical exponents:
\begin{equation}
(\theta_1,\theta_2,\theta_3,\theta_4)\approx (1.04,-0.93,-1.03, -1.78)\,.
\end{equation}

Our expectation about the role of the disconnected interactions seems to be disappointed. No significant improvement is observed, for the critical exponent the value is essentially the same as for LPA $k=8$ and $k=10$ truncations, and the value of the coupling $u_{4*}$ is the same as for the LPA. The only change is for the second critical exponent, whose value is slightly diminished from their purely local version. This, once again, is in complete disagreement with the results of \cite{Eichhorn:2013isa}, and seems to indicate a rapid convergence toward $\theta\approx 1.04$. This intuition is confirmed taking into account triple-trace interactions. We get once again the same fixed point, with $v_{222}=0$ and the same critical exponent $\theta \approx 1.04$. This seems to indicate that triple-trace interactions start to be relevant for $k=10$ truncation, exactly as the double start to be relevant for $k=8$.

\section{Compatibility with Ward identities} \label{sec4}

As we explained in the previous section, there is no preferred notion of scale for the initial model (i.e. for the model without regulator). More precisely, there are canonical notions of deep UV and deep IR: the deep UV being related to the classical action $S(\phi)$, without integration over statistical fluctuations and the opposite deep IR scale, related to the effective action $\Gamma[\Phi]$, when all fluctuations are integrated out. However, there is no canonical way to reach the deep IR region from the deep UV one. All the fluctuations play the same roles, and are indistinguishable ‘‘UV" or ‘‘IR". All the ways that we think to cut through scales are \textit{a priori} allowed and this difficulty is related to the triviality of the Gaussian term. For standard field theories, this is the spectrum of the kinetic operator which provides a canonical path from UV to IR, allowing to classify the fluctuations following their respective energy. But for matrix models, due to the $U(\Lambda)$ invariance, all the eigenvalues of the kinetic operator are the same, and the fluctuations become indistinguishable. This highlights the role of the regulator. The regulator that we introduced broke the global $U(\Lambda)$--invariance at the kinetic level, providing a preferred path from UV to IR and an ordering for partial integrations over quantum fluctuations. \\

The Ward identity that we derived in section \ref{sec2} is a consequence of this symmetry breaking. It arises from the nontrivial variation of the kinetic term under infinitesimal unitary transformation, like the flow equations arise from a nontrivial variation of the kinetic term under a change of the running scale $N$. Both are consequences of the symmetry breaking and have to be treated on the same footing, as nontrivial relations between effective vertices $\Gamma^{(n+2)}$ and $\Gamma^{(n)}$. More precisely, and as it will be clearer in the rest of this section, one can say that the RG equation dictates how to move through increasing scales (from large to small $N$) whereas the Ward identity dictates how to move in the momentum space. As we will see, because of the symmetry breaking, nonlocal derivative-like interactions such that \eqref{nonlocal} appear even in the strictly local sector, and play an important role in the behavior of the RG flow, especially around the UV fixed point.

\subsection{Explicit Ward identities and enlarged theory space}

\textit{i.) Explicit Ward identities. } Taking the derivative with respect to $\partial^2/\partial J_{de}\partial J_{d^\prime e^\prime}$ of the Ward identity \eqref{Ward} and setting $J=0$, we get:
\begin{align}
\nonumber Z_N\,G^{(4)}_{N, de,d^\prime e^\prime, ba, cb} \, [f(a/N,b/N)&-f(c/N,b/N)]\\\nonumber =
\delta_{da}\delta_{eb} G^{(2)}_{N,d^\prime e^\prime,cb}&+\delta_{d^\prime a}\delta_{e^\prime b} G^{(2)}_{N,de,cb}\\
-\delta_{db}\delta_{ec} G^{(2)}_{N, d^\prime e^\prime,ba}&-\delta_{d^\prime b}\delta_{e^\prime c} G^{(2)}_{N, de,ba}\,. \label{equation92}
\end{align}
Setting $d=a$ and $d^\prime=e^\prime=e=c$, and $a\neq c$, we get:
\begin{align}
\nonumber G^{(4)}_{N, ac,cc,cb, ba}& \, [f(a/N,b/N)-f(c/N,b/N)]\\
&\qquad \quad=G^{(2)}_{N,cc,cc}- G^{(2)}_{N, ac,ca}\,,
\end{align}
where we used of the fact that $G^{(4)}_{N, ac,cc,cb, ba}$ must be symmetric under exchange of any pair of indices ($A_i:=(a_ib_i)$):
\begin{equation}
G^{(4)}_{N, A_1A_2A_3A_4}=G^{(4)}_{N, A_{\pi(1)},A_{\pi(2)},A_{\pi(3)}, A_{\pi(4)}}\,,
\end{equation}
for any permutation $\pi$ of four elements. In order to make contact with the parametrization \eqref{localpara}, we decompose $G^{(4)}_{N}$ into connected parts:
\begin{align}
\nonumber &G^{(4)}_{N, A_1A_2A_3A_4}=G^{(4,c)}_{N, A_1A_2A_3A_4}+\bigg(G^{(2)}_{N,A_1A_2}G^{(2)}_{N,A_3A_4}\\
&\qquad +G^{(2)}_{N,A_1A_3}G^{(2)}_{N,A_2A_4}+G^{(2)}_{N,A_1A_4}G^{(2)}_{N,A_3A_2} \bigg)\,. \label{G41}
\end{align}
Discarding all the external propagators of the connected $4$-point function $G^{(4,c)}_{N, A_1A_2A_3A_4}$, we must have:
\begin{equation}
G^{(4,c)}_{N, A_1A_2A_3A_4}=: -\left( \prod_{i=1}^4 g^{(2)}_{A_i}\right) \Gamma^{(4)}_{N,A_1,A_2,A_3,A_4}\,
\end{equation}
where for convenience we introduced the reduced $2$-point components $g^{(2)}_{ab}$, defined from the $2$-point function $G^{(2)}_{ab,cd}$ as:
\begin{equation}
G^{(2)}_{ab,cd}=: g^{(2)}_{ab} \,\delta_{ad}\delta_{bc}\,.
\end{equation}
Discarding multitrace effective interactions, the $4$-point function $\Gamma^{(4)}_{N,A_1,A_2,A_3,A_4}$ has to inherit of the boundary structure of the $4$-point vertices, reflecting the indices conservation along external faces. More precisely, $\Gamma^{(4)}_{N,A_1,A_2,A_3,A_4}$ is assumed to be a sum over the $4!$ permutations for the external pairs $A_i$,
\begin{equation}
\Gamma^{(4)}_{N,A_1,A_2,A_3,A_4}=: \sum_{\pi} \gamma^{(4)}_{A_{\pi(1)},A_{\pi(2)},A_{\pi(3)}, A_{\pi(4)}}\,, \label{G42}
\end{equation}
the boundary structure of $\gamma^{(4)}_{N,A_{\pi(1)},A_{\pi(2)},A_{\pi(3)}, A_{\pi(4)}}$ being fixed, and the relations between external the momenta indices $(a_i,b_i)$ are given by the original $4$-point structure, namely:
\begin{equation}
\gamma^{(4)}_{A_{\pi(1)},A_{\pi(2)},A_{\pi(3)}, A_{\pi(4)}}=:f_{a_1,a_2,a_3,a_4}\delta_{b_1a_2}\delta_{b_2a_3}\delta_{b_3a_4}\delta_{b_4a_1}\,. \label{def99}
\end{equation}
Now, return on equation \eqref{equation92}, and consider the pair $(ac)$. If $a\neq c$, we see that there are four different ways to choose the position of the pair, but for each choice, there remains only one way to fix the relative position of the three other pairs\footnote{We recall that we work in the LPA, then, we consider only connected effective components for effective vertices. We will address the problem of nonconnected boundaries in the next section. }. Therefore \eqref{equation92} may be rewritten as:
\begin{align}
\nonumber 4&Z_N g_{ac}^{(2)}g_{cc}^{(2)}\gamma^{(4)}_{ac,cc,bc,ba} g_{bc}^{(2)}g_{ba}^{(2)}\left[f\left(\frac{a}{N},\frac{b}{N}\right)-f\left(\frac{c}{N},\frac{b}{N}\right)\right]\\
&-Z_Ng_{ac}^{(2)}g_{cc}^{(2)}\left[f\left(\frac{a}{N},\frac{c}{N}\right)-f\left(\frac{c}{N},\frac{c}{N}\right)\right]=-g_{cc}^{(2)}+g_{ac}^{(2)}\,. \label{variation}
\end{align}
Dividing by $g_{ac}^{(2)}g_{cc}^{(2)}$, we deduce the following statement:

\begin{lemma}\label{lemma1}
In the large $N$ limit, the $4$ and $2$-point functions must satisfy the nontrivial relation :
\begin{align}
\nonumber 4&Z_N \gamma^{(4)}_{ac,cc,bc,ba} g_{bc}^{(2)}g_{ba}^{(2)}\left[f\left(\frac{a}{N},\frac{b}{N}\right)-f\left(\frac{c}{N},\frac{b}{N}\right)\right]\\\nonumber
&-Z_N\left[f\left(\frac{a}{N},\frac{c}{N}\right)-f\left(\frac{c}{N},\frac{c}{N}\right)\right]=(g_{cc}^{(2)})^{-1}-(g_{ac}^{(2)})^{-1}\,.
\end{align}
\end{lemma}
\noindent
As explained in the section \ref{sec2}, the Ward identity dictate how to move into the momentum space from ultralocality, whereas the flow equations \eqref{Wett} dictates how to move through scales, from UV to IR. This is in this way that flow equations and Ward identities, both consequences of the $U(\Lambda)$ symmetry breaking, cannot be considered separately. \\

\noindent
Setting $a=c+1$, and for sufficiently large $N$, the difference $f\left(\frac{a}{N},\frac{b}{N}\right)-f\left(\frac{c}{N},\frac{b}{N}\right)$ may be estimated from the same continuous approximation used to compute the sums in the previous section, that is to say:
\begin{equation}
f\left(\frac{a}{N},\frac{b}{N}\right)-f\left(\frac{c}{N},\frac{b}{N}\right)\approx \frac{1}{N} \frac{d}{dx} f\left(x,\frac{b}{N}\right)\bigg\vert_{x=\frac{c}{N}}\,.\label{finitediff}
\end{equation}
Note that this approximation has to be used carefully, and for formal derivations, we may use derivative first as a notation. Computing the derivative for the Litim regulator, we get:
\begin{equation}
\frac{d}{dx} f\left(x,\frac{b}{N}\right)\bigg\vert_{x=\frac{c}{N}}= -2\left(\frac{N}{c+b}\right)^2 \,\Theta\left(1-\frac{c+b}{2N}\right)\,.\label{equ101}
\end{equation}
In the same way, assuming that $g^{(2)}_{ab}$ may be continued as an analytic function $g^{(2)}(x,y)$ for the continuous variables $x,y:=a/N,b/N$, we get:
\begin{equation}
(g_{ac}^{(2)})^{-1}-(g_{cc}^{(2)})^{-1}=\frac{1}{N}\,\frac{d}{dx} g\left(x,\frac{c}{N}\right)\bigg\vert_{x=\frac{c}{N}}+\mathcal{O}\left(\frac{1}{N^2}\right)\label{eq102}
\end{equation}
From equation \eqref{equ101}, it is clear that the windows of momenta allowed in the sum over $b$ from $df/dx:=f^\prime$ is the same as the one allowed by $\dot{r}_N$ in the flow equation \eqref{Wett}. Therefore, the same approximations used to solve the RG equations may be used for $g^{(2)}_{bc}$, $g^{(2)}_{ba}$ and $\gamma^{(4)}_{ac,cc,bc,ba}$. The same situation has been observed for tensor field theory (see \cite{Lahoche:2019vzy}), for several choices of regulator function. Then, one expects that this is not a well consequence of the Litim regulator, but a general feature that the allowed windows of momenta for $\dot{r}_N$ cover the one of $f^\prime$. Moreover, equation \eqref{eq102} points out the existence of a strong relation between $4$-point functions and the momenta variations of the $2$-point functions along the path from the deep UV sector to the IR sector. Therefore, and as we will see explicitly, even in the large $N$ limit, nonlocal interactions such that \eqref{nonlocal} survive at the leading order in $1/N$ and cannot be discarded from any relevant parametrization of the phase space. This argument shows that strictly local potential approximation have to be enlarged with derivative-like interaction to become compatible with Ward identity. As a first improvement, we can consider the following minimal enlargement :
\begin{equation}
\Gamma[\Phi]=\gamma\,\sum_{a,b} \frac{a+b}{2N} \Phi_{ab}\Phi_{ba} +U_N[\Phi]\,, \label{improvedLPA}
\end{equation}
where $U_N[\Phi]$ expands as a single trace like in equation \eqref{potential}. We call improved LPA this parametrization allowing a small deviation from the crude LPA. From this approximation,
\begin{equation}
(g_{ab}^{(2)})^{-1}=Z_N+\gamma\,\frac{a+b}{N}+\mathcal{O}\left(a^2,b^2\right)+Z_Nf\left(\frac{a}{N},\frac{b}{N} \right) \,, \label{derivativeexp}
\end{equation}
and
\begin{equation}
\gamma^{(4)}_{ac,cc,bc,ba}\to\frac{g_4}{4}=(Z_N)^2N^{-1} \frac{u_4}{4}\,.
\end{equation}
Inserting these relations into the lemma \ref{lemma1}, we see that the second term on the left-hand side is exactly compensated with the same term on the right-hand side. Then, setting $c=0$, we get the following statement:
\begin{proposition} \label{propward1}
Up to $1/N$ corrections, and in the improved local potential approximation, the $2$-point derivative coupling $\gamma$ and the local $4$-point coupling $u_4$ satisfy:
\begin{equation}
u_4 \,\bar{\mathcal{L}}_N = -\bar{\gamma}\,, \label{firstWard}
\end{equation}
where $Z_N\bar{\gamma}=:\gamma$ and ${\mathcal{L}}_N=(Z_N)^{-2}N\bar{\mathcal{L}}_N$
\begin{equation}
\mathcal{L}_N:= \sum_b \,(g_{bc}^{(2)})^2 f^\prime (c,b) \big\vert_{c=0}\,.
\end{equation}
\end{proposition}
With the truncation \eqref{improvedLPA}, $\bar{\mathcal{L}}_N$ depends only on $\bar{\gamma}$. Therefore, deriving equation \eqref{firstWard} with respect to $t$ leads to
\begin{equation}
-\beta_4 \bar{\gamma} +u_4(1+u_4 \bar{\mathcal{L}}^\prime_N)\dot{\bar{\gamma}}=0\,. \label{Wardconst}
\end{equation}
This equation that we call \textit{Ward constraint} relies on two beta functions along the history of the RG flow, since $N$ remain large. As an important consequence:
\begin{corollary}
In the large $N$ limit, any fixed point of the flow equations satisfies the Ward constraint \eqref{firstWard}.
\end{corollary}

\noindent
The flow of the nonlocal kinetic coupling $\gamma$ receives two kinds of contributions. A first contribution arises from the derivative with respect to one external momentum of the loop integrals, but a direct computation shows that these variations vanish identically. A second contribution arises from the derivative of the effective vertex themselves. In the local potential approximation, the vertex does not depend on the external momenta. But from the Ward identity, it follows that the ultralocal information determines completely the first derivative with respect to the external momenta, like ultralocal $4$-point coupling $u_4$ determines $\gamma$ in lemma \ref{lemma1}. In order to obtain the first derivative of the $4$-point function, we need to the Ward identity involving $6$-point functions (i.e. derived from \eqref{Ward} deriving four time with respect to the source $J$.). It is more convenient to write the original Ward identity \eqref{Ward} as:
\begin{align}
\nonumber \bigg([r_N(a,b) ]_{ab;de} G^{(2)}_{de,cb}-[r_N(c,b) ]_{bc;de} G^{(2)}_{de,ba}\bigg)\\
- \left(J_{ab} \Phi_{cb}-J_{bc} \Phi_{ba}\right)=0\,.\label{Ward2}
\end{align}
Taking the derivative four time with respect to the classical field $\partial^4/\partial \Phi_{de}\partial \Phi_{d^\prime e^\prime}\partial \Phi_{pq}\partial \Phi_{p^\prime q^\prime}$ of the equation \eqref{Ward2}, we get:
\begin{align*}
&Z_N\bigg[ f\left(\frac{a}{N},\frac{b}{N}\right)-f\left(\frac{c}{N},\frac{b}{N}\right) \bigg] \bigg( g^{(2)}_{ba}g^{(2)}_{cb} \Gamma^{(6)}_{ab,bc,de,d^\prime e^\prime,pq,p^\prime q^\prime }\\
&\quad -6 g^{(2)}_{ba} g^{(2)}_{cb}g^{(2)}_{c^\prime b^\prime} \Gamma^{(4)}_{bc,c^\prime b^\prime,pq,p^\prime q^\prime} \Gamma^{(4)}_{b^{\prime} c^{\prime},ab,de,d^\prime e^\prime} \bigg)=\\
&-\Gamma^{(4)}_{ap^\prime,de,d^\prime e^\prime,pq}\delta_{q^\prime c} -\Gamma^{(4)}_{ap,de,d^\prime e^\prime,p^\prime q^\prime}\delta_{q c}
-\Gamma^{(4)}_{ad^\prime,de,pq,p^\prime q^\prime}\delta_{e^\prime c}\\
&-\Gamma^{(4)}_{ad,d^\prime e^\prime,pq,p^\prime q^\prime}\delta_{ce}
+ \Gamma_{q^\prime c,de,d^\prime e^\prime,pq}^{(4)} \delta_{ap^\prime}
+ \Gamma_{qc,de,d^\prime e^\prime,p^\prime q^\prime}^{(4)} \delta_{ap}\\
&+\Gamma_{e^\prime c,de,pq,p^\prime q^\prime}^{(4)} \delta_{ad^\prime}
+\Gamma^{(4)}_{ec,d^\prime e^\prime,pq,p^\prime q^\prime} \delta_{ad} \,,
\end{align*}
where the $6$ on the second term in the left-hand side is a short notation for the $3\times 2$ terms corresponding to the different pairing of the derived variables. Setting $d^\prime = a$ and $d=e=e^\prime = p=q=q^\prime=p^\prime =c$ for $c\neq a$; and keeping only the leading order contractions in the large $N$ limit, the previous relation reduces to the following lemma:

\begin{lemma} \label{lemmasixpoint}
At the leading order in the $1/N$ expansion, the $6$, $4$ and $2$-point vertex functions must satisfy the following relation:
\begin{align*}
&Z_N\bigg[ f\left(\frac{a}{N},\frac{b}{N}\right)-f\left(\frac{c}{N},\frac{b}{N}\right) \bigg] \bigg( g^{(2)}_{ba}g^{(2)}_{cb} \Gamma^{(6)}_{ab,bc,cc,c a,cc,cc }\\
& -6 g^{(2)}_{ba} g^{(2)}_{cb}g^{(2)}_{c^\prime b^\prime} \Gamma^{(4)}_{c^\prime b^\prime,bc,cc,cc} \Gamma^{(4)}_{b^{\prime} c^{\prime},ab,cc,ca} \bigg)\\
&\qquad =-\left(3\Gamma^{(4)}_{ca,ac,cc,cc}- \Gamma_{cc,cc,cc,cc}^{(4)} \right) \,.
\end{align*}
\end{lemma}
Setting $a=c+1$, and keeping only the first term in the $1/N$ expansion of the difference $f(a/N,b/N)-f(c/N,b/N)$, the argument used for the previous explicit Ward identity holds : the windows of momenta allowed by the distribution $f^\prime(c/N,b/N)$ are the same as for $\dot{r}_N$ involved in the flow equation \eqref{Wett}; and to make sense, the same approximations used to solve this one have to be used in the computation of the Ward identities. Like for the $4$-point vertices we introduce $\gamma^{(6)}$, with fixed boundaries, such that:
\begin{equation}
\Gamma^{(6,c)}_{A_1,A_2,A_3,A_4,A_5,A_6}=\sum_\pi \gamma^{(6)}_{A_{\pi(1)},A_{\pi(2)},A_{\pi(3)},A_{\pi(4)},A_{\pi(5)},A_{\pi(6)}}\,,
\end{equation}
the sum over $\pi$ running through the permutation of six elements, and on the left-hand side, into the sum over $b$, we replace $f_{c,c,c,b}$ by $g_4/4$ and
\begin{equation}
\gamma^{(6)}_{cc,cc,cc,cc,bc,cb} \to \frac{g_6}{6}\,.
\end{equation}
On the right-hand side, from definition \eqref{def99}, $\Gamma_{c,c,c,c}^{(4)}=4! f_{c,c,c,c}$. For $\Gamma^{(4)}_{ac,ca,cc,cc}$ however, there are only $4\times 2$ different configurations for the external indices providing a nonzero contribution. As a result:
\begin{equation}
3\Gamma^{(4)}_{ca,ac,cc,cc}- \Gamma_{cc,cc,cc,cc}^{(4)} = 4! (f_{a,c,c,c}-f_{c,c,c,c})\,.
\end{equation}

As for the $2$-point function, we assume that in the large $N$ limit $f_{c,c,a,c}$ behaves like a continuous function $\tilde{f}(x,c/N)$ for the continuous variable $x=a/N$, such that $\tilde{f}(a/N,c/N)\equiv f_{c,c,a,c}$. We then define, at leading order in $1/N$:
\begin{equation}
f_{c,c,a,c}=: f_{c,c,c,c}+\frac{1}{N}\frac{d\tilde{f}}{dx}\left(x,\frac{c}{N}\right)\bigg\vert_{x=\frac{c}{N}}\,, \label{equation tilde}
\end{equation}
and at the first order in $1/N$, the lemma \ref{lemmasixpoint} becomes, setting $c=0$ and simplifying the global factor $1/N$:
\begin{align}
12\,g_6\mathcal{L}_N -12 g_4^2 \mathcal{U}_N+4!\,\frac{d\tilde{f}}{dx}\left(x,0\right)=0\,,
\end{align}
where we defined:
\begin{equation}
\mathcal{U}_N:=\sum_b \,(g_{bc}^{(2)})^3 f^\prime (c,b) \big\vert_{c=0} \,,\quad \mathcal{U}_N=:(Z_N)^{-3} N \bar{ \mathcal{U}}_N
\end{equation}
From the definition $u_6=(Z_N)^{-3}N^2 g_6$, we finally deduce the following statement, between renormalized quantities:
\begin{proposition}
Up to $1/N$ corrections, and in the improved local potential approximation, the $4$-point derivative coupling and the local $4$ and $6$-point renormalized couplings $u_4$ and $u_6$ are related as:
\begin{align}
2( \,u_6\bar{\mathcal{L}}_N - u_4^2\, \bar{\mathcal{U}}_N)+\,\Xi=0\,, \label{cc}
\end{align}
where we defined:
\begin{equation}
\frac{d\tilde{f}}{dx}\left(x,0\right)=: 4(Z_N)^2 N^{-1}\Xi\,.
\end{equation}
\end{proposition}

\noindent
Equations \eqref{firstWard} and \eqref{cc} show explicitly that the strictly local flow strongly violates the Ward identity. This is especially true at the fixed point, where, from equation \eqref{firstWard} we see that $\bar{\gamma}$ and $u_4$ have to be of the same order, indicating that the regulator scheme strongly influences the nature of the theory space. As we will see in the next section, a systematic analysis, including the flow of the derivative couplings seems to confirm this pessimistic forecast, despite the accordance of the resulting critical exponent with the expected value.

\subsection{Strong deviation with local fixed point}

Now, we move onto derivation of the flow equations in the parametrization \eqref{improvedLPA}. A first change concerns the effective propagator \eqref{eqG}, which becomes:
\begin{align}
(G_N)_{ab,cd}= Z_N^{-1} g_{ba,cd}\, \frac{a+b}{2N} \left( \frac{1}{1+2\bar{\gamma} \left(\frac{a+b}{2N}\right)^2}\right)\,, \label{eqG2}
\end{align}
such that the integral $I_{a}^{(p)}$, equation \eqref{Ip} becomes:
\begin{equation}
I_{a}^{(p)}\approx 2Z_N^{1-p} N \int_{a/2N}^{1} dx\,x^{p-1} \, \frac{1+\eta_N(1-x)}{\left(1+2\bar{\gamma} x^2\right)^p}\,. \label{Ip2}
\end{equation}
To simplify the discussion, we introduce the sequence $\iota_{p,q}(y)$ for the continuous variable $y=a/2N$ such that:
\begin{equation}
\iota_{p,q}(y):=\int_y^1 dx \frac{x^q}{\left(1+2\bar{\gamma} x^2\right)^p}\,,
\end{equation}
and:
\begin{equation}
I_{a}^{(p)}\approx 2Z_N^{1-p} N\left[\iota_{p,p-1}(y)+\eta_N\left(\iota_{p,p-1}(y)-\iota_{p,p}(y)\right) \right]\,.
\end{equation}
In addition we defined the renormalized loop $\bar{I}_{a}^{(p)}:=Z_N^{p-1} N^{-1} I_{a}^{(p)}$.
Equation \eqref{eqZ} is then transformed as:
\begin{equation}
\dot{Z}_N=-\frac{2Ng_{4,N}}{Z_N}\left[\iota_{2,1}(0)+\eta_N\left(\iota_{2,1}(0)-\iota_{2,2}(0)\right) \right]\,,\label{eqZ2}
\end{equation}
solved as:
\begin{equation}\label{122}
\eta_N=-\frac{2u_4\,\iota_{2,1}}{1+2u_4\left(\iota_{2,1}-\iota_{2,2}\right) }\,,
\end{equation}
where we used  the concise notation $\iota_{p,q}\equiv \iota_{p,q}(0)$. In the same way, we get for $\beta_4$, in replacement of \eqref{eqbeta4}:
\begin{align}
 \beta_4=(1-2\eta) u_{4}& + 8 u_{4}^2\left[ \iota_{3,2} +\eta\left(\iota_{3,2}-\iota_{3,3}\right) \right]\\
&-4u_{6}\left[ \iota_{2,1} +\eta\left(\iota_{2,1}-\iota_{2,2}\right) \right]\,. \label{eqbeta42}
\end{align}
The flow equation for $\gamma$ can be deduced from \eqref{equationwettexp1}, like $\eta_N$. From definition:
\begin{equation}
\frac{\gamma}{N}\equiv \frac{d}{da} \Gamma^{(2)}_{ab,ba}\big\vert_{a=b=0} \,,
\end{equation}
we get ($\beta_\gamma\equiv \dot{\bar{\gamma}}$):
\begin{align}
\nonumber \beta_\gamma=-\eta\, \bar{\gamma} -8 u_4& \left[\iota_{2,1}^\prime+\eta\left(\iota_{2,1}^\prime-\iota_{2,2}^\prime \right) \right]\\
&\qquad -12 \Xi \left[\iota_{2,1}+\eta\left(\iota_{2,1}-\iota_{2,2}\right) \right]\,.
\end{align}
It is easy to check that the involved derivatives $\iota_{p,q}^\prime\equiv \iota_{p,q}^\prime(0)$ vanish identically
\begin{equation}
\iota_{p,q}^\prime= -\frac{x^q}{\left(1+2\bar{\gamma} x^2\right)^p} \bigg\vert_{x=0}=0\,,\quad \forall q\neq 0\,,
\end{equation}
such that the equation for $\dot{\bar{\gamma}}$ reduces to:
\begin{align}
\nonumber \beta_\gamma&=-\eta\, \bar{\gamma} -12 \Xi \left[\iota_{2,1}+\eta\left(\iota_{2,1}-\iota_{2,2}\right) \right] \\
&=-\eta\, \bar{\gamma} +24 \left(u_6 \iota_{2,0}-u_4^2\iota_{3,1}\right) \left[\iota_{2,1}+\eta\left(\iota_{2,1}-\iota_{2,2}\right) \right]\,,
\label{eqgamma1}
\end{align}
where we took into account that what we called $\bar{\mathcal{L}}_N$ and $\bar{\mathcal{U}}_N$ may be expressed in terms of the sequences $\iota_{p,q}$,
\begin{equation}
\bar{\mathcal{L}}_N:= - \iota_{2,0}\,,\qquad \bar{\mathcal{U}}_N:=- \iota_{3,1}\,. \label{LN}
\end{equation}
Note that the origin of the factor $12\equiv 4\times 3$ in front of $\Xi$ in equations \eqref{eqgamma1} counts the different localizations for the derivative $\tilde{f}^\prime$ (see equation \eqref{equation tilde}) on the vertex itself, as pictured on Figure \ref{confXi} below. \\

\begin{figure}
\includegraphics[scale=0.5]{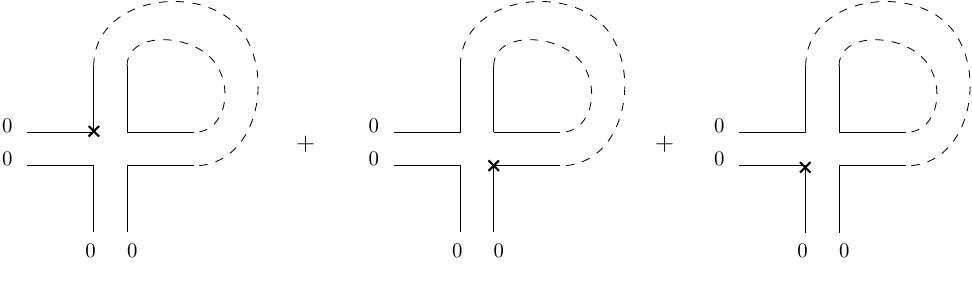}
\caption{The three contributions to the derivative of the effective vertex, the cross means location of the derivative $\tilde{f}^\prime$.}. \label{confXi}
\end{figure}

Another expression for $\dot{\gamma}$ comes from the Ward identity, equation \eqref{Wardconst}, namely
\begin{equation}
\beta_\gamma=-\frac{ \bar{\mathcal{L}}_N}{1+u_4 \bar{\mathcal{L}}^\prime_N} \beta_4\,.
\end{equation}
Obviously $\iota^\prime_{2,0}=-4\iota_{3,2}$ (where the {\it prime} means the derivative with respect to $\bar\gamma$), so that the equation for $\dot{\gamma}$ reduces to:
\begin{equation}
\beta_\gamma=\frac{\iota_{2,0}}{1+4u_4 \iota_{3,2}} \beta_4\,. \label{WC}
\end{equation}
From $\beta_4$ given by equation \eqref{eqbeta42}, and equation \eqref{eqgamma1}, we can deduce $u_6$ in terms of $u_4$ and $\bar{\gamma}$ dynamically along the RG flow, $u_6=f(u_4,\bar{\gamma})$, with
\begin{align*}
&f(u_4,\bar{\gamma})\\
&\quad =-\frac{(1-\eta)\bar{\gamma}+2u_4^2\left(2\iota_{2,0} \bar{I}_0^{(3)}-3\iota_{3,1} \bar{I}_0^{(2)}(1+4u_4\iota_{3,2})\right)}{\iota_{2,0} \bar{I}_0^{(2)}[7+24u_4\iota_{3,2}]}\,,
\end{align*}
which, from the Ward constraint \eqref{propward1} can be translated as a function on $\bar{\gamma}$ only. At this level of approximation, the problem is then completely closed. The two parameters $\bar{\gamma}$ and $u_4$ fix $u_6$, which fix $u_8$ and so one. This conclusion highlights two points. First, the role played by the derivative couplings, second that the improved local potential parametrization \eqref{improvedLPA}, which involve an infinite number of couplings can be, in fact, reduced to a one-dimensional manifold. Obviously, enlarging the theory space with more derivative and/or disconnected interactions, we lost this property. Moreover, note that we neglected the flow of the derivative coupling $\dot{\Xi}\approx 0$. \\

We now move on the essential motivation to build the improvement discussed in the previous paragraph: the investigation of the global fixed point solutions of the flow equations. Our strategy is the following. Setting $\beta_4=0$, from the linearity of the equation in the sixtic coupling, we fix $u_6$ \textit{uniquely} in terms of $u_4$ and $\bar{\gamma}$ through a relation of the form $u_6=F(u_4,\bar{\gamma})$. Moreover, from the first Ward identity given by Proposition \eqref{propward1}, $u_4$ and $\bar{\gamma}$ are not independent, $u_4= \bar{\gamma}/\iota_{0,2}$, therefore :
\begin{equation}
u_6=F\left(\frac{\bar{\gamma}}{\iota_{0,2}},\bar{\gamma}\right)\,.
\end{equation}
Explicitly:
\begin{equation}
F(u_4,\bar{\gamma})=\frac{(1-2\eta) u_{4}+ 8 u_{4}^2\left[ \iota_{3,2} +\eta\left(\iota_{3,2}-\iota_{3,3}\right) \right]}{4\left[ \iota_{2,1} +\eta\left(\iota_{2,1}-\iota_{2,2}\right) \right]}\,.
\end{equation}
Inserting these relation into equation \eqref{eqgamma1}, and setting $\dot{\bar{\gamma}}=0$, we deduce  the following:
\begin{proposition} \label{fixedpointequation}
In the large $N$ limit, all the fixed points of the improved LPA have to be solution of the following equation:
\begin{align*}
0=-\beta_\gamma\equiv \eta\, \bar{\gamma} -&24 \bigg(F\left(\frac{\bar{\gamma}}{\iota_{2,0}},\bar{\gamma}\right) \iota_{2,0}-\left(\frac{\bar{\gamma}}{\iota_{2,0}}\right)^2\iota_{3,1}\bigg) \bar{I}_0^{(2)}\,.
\end{align*}
\end{proposition}
This equation can be solved numerically. One may expect that the dynamical definition of $u_6$, $u_6=f(u_4,\bar{\gamma})$ breaks down at the fixed point, because both $\beta_\gamma$ and $\beta_4$ vanish at this point. It is not hard however to show that:

\begin{lemma}
The effective RG flow, described by the function $f(u_4,\bar{\gamma})$ satisfies:
\begin{equation}
f(u_4,\bar{\gamma})\big\vert_{\bar{\gamma}^*}=F(u_4,\bar{\gamma})\vert_{\bar{\gamma}^*}\,,
\end{equation}
ensuring continuity at the fixed point.
\end{lemma}

\noindent
\textit{Proof.} The proof is elementary. Let us rewrite our set of flow equations as:
\begin{align}
\beta_\gamma&=-\eta \bar{\gamma}-6(a_\gamma u_6-b_\gamma u_4^2) \\
\beta_4&=(1-2\eta) u_4+b_4 u_4^2-a_4 u_6\,,
\end{align}
and the relation between them coming from Ward identity as $\beta_\gamma = A \beta_4$. Using the last one, we get the explicit expression for $f$:
\begin{equation}
u_6^\sharp= \frac{(1-2\eta) Au_4+\eta \gamma+ (b_4A-6b_\gamma)u_4^2)}{a_4 A-6a_\gamma}\,.
\end{equation}
Moreover, setting $\beta_\gamma=0$ on one hand, and $\beta_4=0$ on the second hand; we get respectively the two solutions:
\begin{align}
u_6^*&=\frac{(1-2\eta)u_4+b_4 u_4^2 }{a_4}\,,\\
u_6^{**}&=\frac{-\eta \bar{\gamma}+6b_\gamma u_4^2}{6a_\gamma}\,.
\end{align}
Inserting these two solutions into the expression of the dynamical coupling $u_6$, we get:
\begin{equation}
u_6^\sharp = \frac{A a_4 u_6^*-6a_\gamma u^{**}_6}{A a_4-6a_\gamma}\,.
\end{equation}
For a global fixed point $u_6^*=u_6^{**}$. Therefore, without singularity of the involved coefficients, we get $u_6^{\sharp}\equiv u_6^*$.
\begin{flushright}
$\square$
\end{flushright}

\begin{figure}
\includegraphics[scale=0.5]{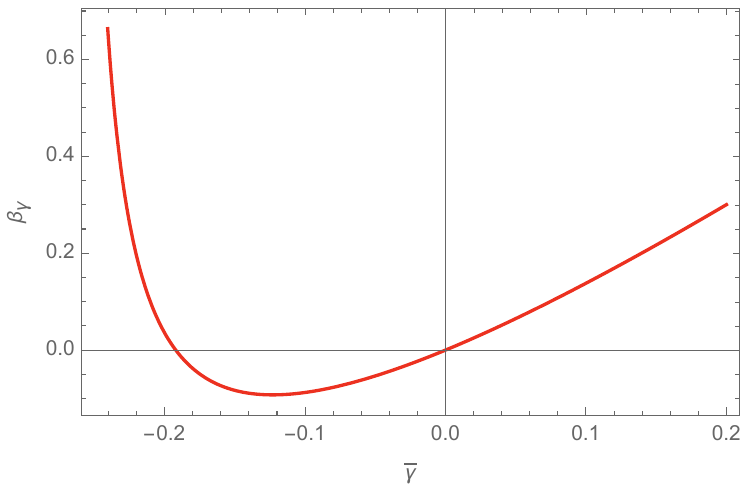}
\caption{Numerical plot of the beta function $\beta_\gamma$ expressed in term of $\bar{\gamma}$ only. The function has two nodes. The first one for $\bar{\gamma}=0$ corresponds to the Gaussian fixed point, the other, for $\bar{\gamma}\approx -0.19$ corresponds to an attractive fixed point.} \label{figplotbetagamma}
\end{figure}

\noindent
The numerical plot of the $\beta$-function $\beta_\gamma$ is provided in Figure \ref{figplotbetagamma}, showing three zeros. The first one, for $\bar{\gamma}=0$ corresponds to the Gaussian fixed point $u_4=0$;
the other, for $\bar{\gamma}\approx -0.19$ is UV-attractive, and seems to be in qualitative agreement with the UV attractive fixed point relevant for double scaling limit. Computing $u_4^*$ from Ward identity, we get:
\begin{equation}
u_4^*=\frac{\bar{\gamma}^*}{\iota_{2,0}(\bar{\gamma}^*)} \approx -0.14\,,\qquad \eta_* \approx 0.23\,,
\end{equation}
The critical exponent can be computed from \eqref{WC}, computing the derivative of $\beta_\gamma$ with respect to $\bar{\gamma}$ at the fixed point. Numerically, we get:
\begin{equation}
\theta\approx 1.53\,.
\end{equation}
We discovered a relevant direction, but in strong disagreements with the theoretical predictions for the local matrix model. This is not necessarily surprising. Indeed, the value of $\bar{\gamma}$ at the fixed point is relatively large, indicating a large deviation from the local model, likely to generate important qualitative differences. Thus, taking into account the Ward identities with an arbitrary regulator generates non negligible effects, making the predictions of the RG for the local model unreliable. However, the approach has the merit of indicating a clear criterion: we expect the predictions to be closer to those of the local model the smaller the value of $\bar{\gamma}$ at the fixed point is. Thus, any regularization minimizing its effects will gain in reliability with respect to investigations in the theoretical space of local models and their critical properties. Let us finally note that the flow of $\bar{\gamma}$ being an effect of the regulator, it also gives an indication concerning the dependence of the results on the choice of the latter. 
\medskip

We considered only the minimal crude truncation in the space of derivative couplings, showing the instability of the ultralocal sector due to the Ward identity. But morally, all the derivative couplings have to  contribute on the left-hand side of the Ward identities, and equation \eqref{Ward2} can be viewed for instance as the minimal truncation of a complete equation, involving an infinite set of couplings. To be more precise, let us introduce the following graphical notation. For each derivative operator like :
\begin{equation}
V[\Phi]_{2,3,1,0}:=\sum_{a,b,c,d} \left(\frac{a}{N}\right)^2\left(\frac{b}{N}\right)^3\left(\frac{c}{N}\right) \Phi_{ab}\Phi_{bc}\Phi_{cd}\Phi_{da}\,,
\end{equation}
we adopt a graphical representation as:
\begin{equation}
V[\Phi]_{2,3,1,0}\equiv \vcenter{\hbox{\includegraphics[scale=1]{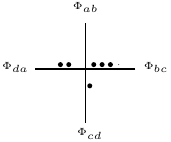} }}\,,
\end{equation}
the dots counting the ‘‘derivative insertions". Keeping only connected interactions, the enlarged theory space then can  include all the possible ‘‘dotted" interactions,
\begin{align}
\nonumber & \Gamma_N[\Phi] \sim \vcenter{\hbox{\includegraphics[scale=1.3]{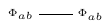} }}+ \vcenter{\hbox{\includegraphics[scale=1.3]{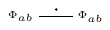} }}
\\\nonumber
&+\vcenter{\hbox{\includegraphics[scale=1.3]{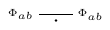} }}+ \vcenter{\hbox{\includegraphics[scale=1.3]{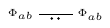} }}+ \vcenter{\hbox{\includegraphics[scale=1.3]{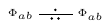} }}\\\nonumber
&+ \vcenter{\hbox{\includegraphics[scale=0.7]{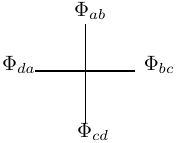} }}
+ \vcenter{\hbox{\includegraphics[scale=0.7]{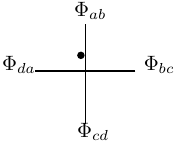} }}
+\vcenter{\hbox{\includegraphics[scale=0.7]{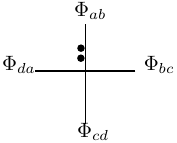} }}\\\nonumber &+\vcenter{\hbox{\includegraphics[scale=0.7]{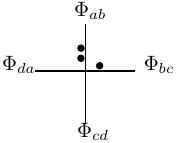} }}+ \vcenter{\hbox{\includegraphics[scale=0.7]{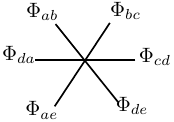} }}+ \vcenter{\hbox{\includegraphics[scale=0.7]{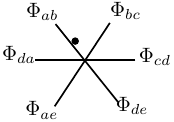} }}\\\nonumber
&+ \vcenter{\hbox{\includegraphics[scale=0.7]{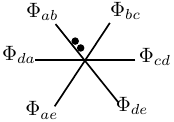} }}+\vcenter{\hbox{\includegraphics[scale=0.7]{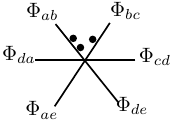} }}+\cdots \,,
\end{align}
Such that with this short notation, the Ward identities \eqref{Ward2} and \eqref{cc} rewrite as:
\begin{equation}
\vcenter{\hbox{\includegraphics[scale=1.5]{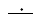} }}=\vcenter{\hbox{\includegraphics[scale=1.3]{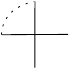} }}\vcenter{\hbox{\includegraphics[scale=1.3]{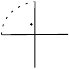} }}\vcenter{\hbox{\includegraphics[scale=1.3]{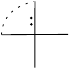} }}\vcenter{\hbox{\includegraphics[scale=1.3]{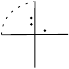} }}\cdots \,,
\end{equation} \label{Ward22}
and:
\begin{align}
\nonumber \vcenter{\hbox{\includegraphics[scale=1]{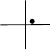}}}&=\vcenter{\hbox{\includegraphics[scale=1]{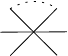}}}+\vcenter{\hbox{\includegraphics[scale=1]{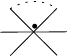}}}+\vcenter{\hbox{\includegraphics[scale=1]{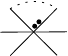}}}+\vcenter{\hbox{\includegraphics[scale=1]{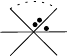}}}\\\nonumber
&+\vcenter{\hbox{\includegraphics[scale=1]{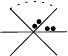}}}+\cdots -\vcenter{\hbox{\includegraphics[scale=1]{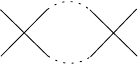}}}\\\nonumber
&-\vcenter{\hbox{\includegraphics[scale=1]{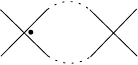}}}
-\vcenter{\hbox{\includegraphics[scale=1]{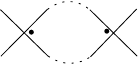}}}\\
&-\vcenter{\hbox{\includegraphics[scale=1]{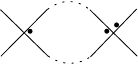}}}
-\vcenter{\hbox{\includegraphics[scale=1]{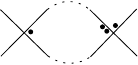}}}-\cdots \,,
\end{align} \label{cc2}
and so one for higher derivative couplings; the dotted edge meaning sums like $\mathcal{L}_N$ and $\mathcal{U}_N$, or higher momenta when dots appear along the resulting closed face. We have then to deal with a proliferating number of derivative couplings. A crude truncation over the theory space, like we considered in this section improves the result with respect to a naive ultralocal truncation. But the result seems not to be satisfactory, because of the strong dependence of the derivative couplings. A way to solve this difficulty could be to investigate the dependence of the fixed point on the choice of the regulator. If the resulting fixed point and its associated critical exponents depend slightly on the choice of the regulator, for a large range of them, this can be a strong argument in favor of the reliability of our result. This is the strategy that we will discuss in the next section.

 Before starting the next section, and without going into the technical details concerning the optimization of the regulator, let us provide here some important comments on the choice of the regulator and its optimization condition. For the well known exact models (we denote by exact model the solvable model with the exact solution on the flow), a regulator of the FRG analysis is said to be optimal if the corresponding fixed points and critical exponents are very close to the exact results. Note also that all the regulators must carefully check the limits $\Gamma_{N\to\Lambda}=S$ (the microscopic action) and $\Gamma_{N\to 0}=\Gamma$ (the effective action). In the case of the nonsolvable model, we do not have rigorous criteria to fix the choice of an optimal regulator because no comparison of the results coming from the FRG can be made with the exact results. In the case of this paper the exact result is well known in the literature \cite{DiFrancesco:1993cyw} and therefore the comparison is possible as well as the choice of the optimal regulator.  Our approach to optimization is different from the general field-theoretical approach to optimization discussed in \cite{Litim:2000ci},\cite{Litim:2001fd}-\cite{Pawlowski:2005xe}. The choice of a truncation and a optimal  regulator will be determined by their agreement with the exact results. Thus, we will deform the Litim regulator in the hope that the deformation parameters will be chosen in order to properly approach the exact results well known in the literature. Finally the Ward identity is also a constraint that can chart the path in choosing such an optimal regulator.

\subsection{Dependency on the regulator: a first look}
In order evaluate qualitatively the dependence of our results on the choice of the regulator, a simple way is to introduce a parametrization depending on a small number of parameters. Following  \cite{Canet:2002gs}, we consider the following parametrization:
\begin{equation}
r_N \to \alpha r_N \,.
\end{equation}
We get the effective propagator into the range $a+b\leq 2N$:
\begin{equation}
(G_N)_{ab,cd}= \, \frac{a+b}{2N} \left( \frac{(\alpha Z_N)^{-1} g_{ba,cd}}{1+\frac{2\bar{\gamma}}{\alpha} \left(\frac{a+b}{2N}\right)^2+\frac{1-\alpha}{\alpha}\frac{a+b}{2N}}\right)\,,
\end{equation}
such that, in the integral approximation, $I_a^{(p)}$ becomes
\bea
I_a^{(p)}=2(\alpha Z_N)^{1-p}N\int_{\frac{a}{2N}}^1\,dx\, \frac{x^{p-1}[\eta_N(1-x)+1]}{\Big(1+\frac{2\bar{\gamma}}{\alpha}x^2+\frac{1-\alpha}{\alpha}x\Big)^p} \,,
\eea
and:
\bea
\iota_{p,q}(y)=\alpha^{1-p}\int_{y}^1dx\,\frac{x^q}{\Big(1+\frac{2\bar\gamma}{\alpha}x^2+\frac{1-\alpha}{\alpha}x\Big)^p}\,.
\eea
From these definitions it is straightforward to get the flow equations as \eqref{122}, \eqref{eqbeta42} and \eqref{eqgamma1}. Note that equations \eqref{LN} and \eqref{WC} are unchanged. Investigating the fixed point structure for several values of $\alpha$, we get the table \ref{tablegras}. \\

As discussed, the fixed point like the critical exponent has a nontrivial dependency concerning the choice of the regulator; that we expect to be a consequence of the role played in the computation of the critical exponents by the derivative coupling $\gamma$. This dependence seems to be very strong, except in two regions, in the vicinity of $\alpha \approx 1$ and $\alpha \approx 2.5$, where critical exponents reach values $\theta \approx 1.5$ and $\theta \approx 1.2$ respectively. Around these two extrema, the dependency with respect to the regulator (in the considered parametrization) is small. Note that, it is possible to fine tune the value of $\alpha$ to reproduce exactly the expected critical exponent. On find $\theta \approx 0.8$ for $\alpha \approx 0.563$. However, in this region of the parameter space, $\theta$ depends strongly on $\alpha$, and the reliability is poor. \\

Another  popular choice for the regulator is the exponential one:
\begin{equation}
r_N(a,b)=\frac{Z_N}{e^{(\frac{a+b}{2N})}-1}\,,
\end{equation}
and numerical investigations leads to the same conclusions as for the Litim's choice: A strong dependency of the relevant quantities on the regulator. 
{\color{red}

\begin{figure}
\begin{tabular}{|c|c|c|c|c|}
\hline
Parameter $\alpha$& Critical exponent $\theta$  &$\eta$ & $u_4$  \\
\hline
0.4&-2.81&0.4&-1.13\\
\hline
0.5 & 0.07&0.36&-0.70 \\
\hline
0.6&1.05&0.33&-0.47\\
\hline
0.7&1.4&0.3&-0.32\\
\hline
0.8&1.53&0.28&-0.23\\
\hline
0.9&1.55&0.25&-0.17\\
\hline
1&1.53&0.24&-0.14\\
\hline
1.1&1.49&0.22&-0.10\\
\hline
1.2&1.45&0.20&-0.08\\
\hline
1.6&1.3&0.15&-0.03\\
\hline
1.8&1.24&0.13&-0.025\\
\hline
2&1.20&0.11&-0.02\\
\hline
2.5&1.18&0.08&-0.008\\
\hline
3&1.22&0.06&-0.004\\
\hline
3.5&1.36&0.05&-0.002\\
\hline
\end{tabular}
\caption{Numerical computation of the fixed point and critical exponent as a function of the regulator dilatation parameter $\alpha$. We can remark that a very small variation of the parameter $\alpha$ drastically modifies the critical exponent $\theta$.}\label{tablegras}
\end{figure}
}

\section{Discussions and conclusion} \label{sec6}

The Ward identities, like \eqref{Ward2} and \eqref{cc} highlight the role played by the regulator in the emergence of the derivative couplings. We can stress a parallel between flow evolution and divergence of the flow toward the derivative sector. In both cases, this is the variation of the propagator -- for $N$ or $a/N\equiv x$  that generates the moving into the theory space, in ‘‘scales" or ‘‘momenta" directions respectively. Moreover, the two transformations are not generally independents. For a regulator of the form:
\begin{equation}
r_N(x,y)=Z_N f(x,y)\,,
\end{equation}
we get:
\begin{equation}
\dot{r}_N(x,y)=\eta_N\,r_N(x,y)-Z_N\left(x\frac{\partial f}{\partial x}+y\frac{\partial f}{\partial y}\right)\,. \label{derivative}
\end{equation}
The first term is intrinsically associated with the RG flow; however, the second part involves derivative for the momenta, which are the generators of the momentum displacements in Ward identities.

\begin{figure}[H]
\includegraphics[scale=0.5]{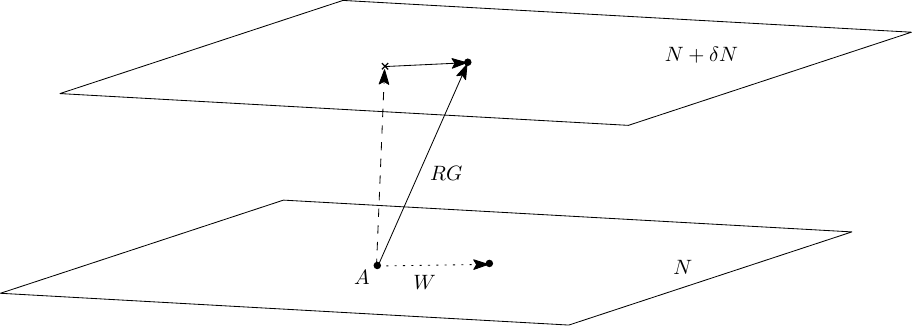}
\caption{Qualitative behavior of the RG map and Ward operator (W) into the full theory space. Starting from a point in the theory space (A), the Ward operator (the dotted arrow) generates horizontal moves at fixed $N$, whereas the RG map has both vertical and horizontal components.}\label{rep}
\end{figure}

Heuristically, one may picture the global dynamics as follows (see Figure \ref{rep}). Starting from a point at scale $N$ in the full theory space, the Ward operator allows moving horizontally, at fixed $N$ toward the derivative interactions world. In the same way, the RG map allows to move vertically, from the scale $N$ to the scale $N+\delta N$, but due to the second terms of the right-hand side of \eqref{derivative}, the RG transformation generates as well a horizontal displacement. This is another way to understand the instability of the local phase space, pointed out in the previous section.
Therefore, and despite the accordance of our results with the expected ones, especially about the value of the critical exponent, and the apparent qualitatively small dependence on the regulator in a small range of values around $\alpha=1$; one cannot conclude that our results have anything to do with the original model, the explored region of the theory space being very far from one of the original ultralocal ones. \\

From the last picture, a question remains open. Can you build a RG map which is the most vertical as possible, at least for $N$ sufficiently large, in such a way that $\mathcal{L}_N$, $\mathcal{U}_N$ and their higher momenta remains small enough, such that derivative couplings can be discarded from the RG flow? This question seems to be very difficult in regard to the complex hierarchical structure of the flow equations that we discussed in this paper. A heuristic attempt to solve this problem, or at least to build a flow which remains vertical for a long time is to choose a regulator such that $\mathcal{L}_N$, $\mathcal{U}_N$ vanish or become small for vanishing $\bar{\gamma}$. This can be achieved for instance with a regulator of the form:
\begin{equation}
f\left(\frac{a}{N},\frac{b}{N}\right):= \left(\frac{2N}{a+b}-1 \right)\Theta\left(\alpha-\frac{a+b}{2N} \right)\,,\label{regulnew}
\end{equation}
where the  parameters $\alpha$  have to be fine-tuned such that $\mathcal{L}_N$ vanish and , $u_4^2\mathcal{U}_N$  becomes small  for $\bar{\gamma}=0$. By solving   $\mathcal{L}_N=0$ we get the solution $\alpha=2$. It is easy to check that this regulator satisfies the four requirements enumerated above equation \eqref{regulator}. The corresponding flow equations can be easily deduced from our previous analysis. The condition $\mathcal{L}_N\vert_{\gamma=0}=0$ and $|\mathcal{U}_N\vert_{\gamma=0}|\approx 1$ allows to keep $\Xi=\bar{\gamma}=0$ with a very good approximation along the flow for a long time, in regards to the rapidity of the convergence of the truncated expansions.  Setting $\bar{\gamma}=0$, the flow equations become the following:
\begin{proposition}
In the large $N$ limit and for the fine-tuned regulator \eqref{regulnew}, the most vertical truncated flow equations in the LPA, up to $\Phi^{10}$-interactions, write as:
\begin{align}
\nonumber \beta_4=&(1-2\eta)u_4+8u_4^2[\iota^{(1)}_3 \eta+\iota^{(2)}_3+\partial \iota_3]\\\nonumber
&-4u_6[\iota^{(1)}_2 \eta+\iota^{(2)}_2+\partial \iota_2]\,,
\end{align}
\begin{align}
\nonumber \beta_6&=(2-3\eta)u_6+24 u_6u_4[\iota^{(1)}_3 \eta+\iota^{(2)}_3+\partial \iota_3]\\\nonumber
&-12 u_4^3[\iota^{(1)}_4 \eta+\iota^{(2)}_4+\partial \iota_4]-6 u_8 [\iota^{(1)}_2 \eta+\iota^{(2)}_2+\partial \iota_2]\,,
\end{align}
\begin{align*}
\beta_8&=(3-4\eta)u_8+16 u_4^4[\iota^{(1)}_5 \eta+\iota^{(2)}_5+\partial \iota_5]\\
&+16 u_6^2 [\iota^{(1)}_3 \eta+\iota^{(2)}_3+\partial \iota_3]-48 u_6 u_4^2 [\iota^{(1)}_4 \eta+\iota^{(2)}_4+\partial \iota_4]\\
&+32 u_8u_4 [\iota^{(1)}_3 \eta+\iota^{(2)}_3+\partial \iota_3]\,,
\end{align*}
where we used of the definitions:
\begin{equation*}
\eta:=-2u_4\frac{\iota^{(2)}_2+\partial\iota_2}{1+2u_4 \iota^{(1)}_2}
\end{equation*}
and:
\begin{equation}
\iota^{(1)}_p:=\iota^{(2)}_p-\iota^{(2)}_{p+1}\,,
\end{equation}
\begin{equation}
\iota^{(2)}_p:=\frac{2^p}{p}\,,
\end{equation}
\begin{equation}
\partial \iota_p:=\alpha^{p}(1-\alpha) \,.
\end{equation}
\end{proposition}

\begin{figure}[H]
\begin{tabular}{|c|c|c|c|c|}
\hline
truncation order $k$ & $\theta_1$ & $\theta_2$ & $\theta_3$ &anomalous dimension $\eta$\\
&&&&
\\
\hline
6&1.02&--&--&0.08\\
\hline
8 & 1.01&-1.19&--&0.05\\
\hline
10&1.00&-1.09&-2.43&0.03\\
\hline
\end{tabular}
\caption{Numerical results for vertical truncations from $k=6$ to $k=10$. We see that increasing the number of interactions does not change the value of the positive critical exponents, the other one corresponding to irrelevant directions. Moreover, the anomalous dimension is very small in comparison to truncation with the standard Litim regulator. }\label{table2}
\end{figure}

Investigating numerically the successive truncations, for $k=6$, $k=8$ and $k=10$ like in the section \ref{sec3}, we get only one fixed point with one relevant direction, the details being summarized in Table \ref{table2} below.  Interestingly, the convergence of the truncations seems to be improved with respect to the ones considered in section \ref{sec3}. We get only one relevant direction, with a critical exponent matching with the perturbative result. Note that no significant improvement arises from the nonperturbative effects. Moreover, the value of the relevant critical exponent seems to be insensitive to the level of the truncation. However, the value of the corresponding coupling is in strong discordance with the expected one. We get a positive and very small value for $u_4$, $u_4\approx 0.016$ for $k=6$, $u_4\approx 0.01$ for $k=8$ and $u_4\approx 0.008$ for $k=10$; the values of the other couplings being of very small magnitude with respect to these values. \\

This  observation, as mentioned, is not presented as a rigorous way to build a solution for nonperturbative RG equation, but as a qualitative illustration of how we can deal with derivative couplings to keep the flow in the purely local sector. A more complete investigation has to be carried out on this subject. Other more sophistical methods, using, for instance, background fields to constrain the flow along the vertical direction are expected to be helpful to realize such an RG map. We keep these investigations to a forthcoming work.


\onecolumngrid

\end{document}